\documentclass[aps,rmp]{revtex4}
\usepackage{natbib}
\usepackage{graphics}
\usepackage{epsfig}
\usepackage{psfig}
\newcommand{\beq}{\begin{equation}}
\newcommand{\dd}{\partial}
\newcommand{\eeq}{\end{equation}}
\newcommand{\bea}{\begin{eqnarray}}
\newcommand{\eea}{\end{eqnarray}}
\newcommand{\lsim}{\stackrel{<}{\scriptstyle \sim}}
\newcommand{\gsim}{\stackrel{>}{\scriptstyle \sim}}

\newcommand{\vf}{\varphi}

\newcommand{\e}{{\cal E}_\omega}

\begin{document}

\begin{flushright}
\begin{tabular}{l}
SCIPP-2003/01 \\
UCLA/03/TEP/08
\end{tabular}
\end{flushright}

\vspace{-5mm}

\title{The Origin of the Matter-Antimatter Asymmetry}

\author{Michael Dine}
\affiliation{Santa Cruz Institute for Particle Physics, Santa Cruz, CA
  95064}
\author{Alexander Kusenko}
\affiliation{
Department of Physics and Astronomy, University of California, Los Angeles,
CA 90095-1547,
}
\affiliation{RIKEN BNL Research Center, Brookhaven National
Laboratory, Upton, NY 11973
}

\begin{abstract}

Although the origin of matter-antimatter asymmetry remains unknown,
continuing advances in theory and improved experimental limits have ruled
out some scenarios for baryogenesis, for example the sphaleron baryogenesis
at the electroweak phase transition in the standard model.  At the same
time, the success of cosmological inflation and the prospects for
discovering supersymmetry at the LHC have put some other models in sharper
focus.  We review the current state of our understanding of baryogenesis
with the emphasis on those scenarios that we consider most plausible.

\end{abstract}

\maketitle
\tableofcontents

\section{Introduction}
\label{sec:intro}

When we observe the universe, the most obvious, and easily
studied, objects are stars and gas, made up of
protons, neutrons and electrons. Astrophysicists speak of the density
of protons and
neutrons, which constitute the bulk of the mass of this matter, as
the baryon content of the universe.

But we know that there is much more to the universe than baryons.
By indirect means, astronomers have established that approximately 1/3 of
the energy density of the universe is in the form of some
non-baryonic matter, referred to as the dark matter, while 2/3 is in a form
with negative pressure, perhaps a cosmological constant.
The baryons make up a mere $5\%$ of the total energy density of the universe.

Another, even more striking measure of the smallness of the baryon
density is provided by the ratio of baryons to photons in
the Cosmic Microwave
Radiation Background (CMBR).  Big Bang nucleosynthesis gives a good
measure of the baryon density; this measurement is well supported
by recent measurements of the fluctuations of the
cosmic microwave radiation background.  As a result, the
ratio of baryons to photons is now known to about $5\%$~\cite{map}:

\beq
{n_{_B} \over n_\gamma} = \left (6.1{\,}^{+0.3}_{-0.2} \right )
\times  10^{-10}.
\label{baryondensity}
\eeq
There is good
evidence that there are no large regions of antimatter at any but
cosmic distance scales~\cite{cohenetal}, although some small domains of
antimatter in the matter-dominated universe are not ruled out by
observations~\cite{anti_1,anti_2,anti_3}.

It was A. Sakharov who first suggested that the baryon density might
not represent some sort of initial condition, but might be
understandable in terms of microphysical laws~\cite{sakharov}.
He listed three ingredients to such an understanding:
\begin{enumerate}
\item  Baryon number violation must occur in the fundamental
laws.  At very early
times, if baryon number violating interactions were in equilibrium,
then the universe can be said to have ``started" with zero baryon
number.   Starting with zero baryon number, baryon number
violating interactions are obviously necessary if the universe is
to end up with a non-zero asymmetry.  As we will see, apart from
the philosophical appeal of these ideas, the success of
inflationary theory suggests that, shortly after the big bang, the
baryon number was essentially zero.
\item  CP-violation:  If CP (the product of charge conjugation
and parity) is conserved, every reaction
which produces a particle will be accompanied by a reaction
which produces its antiparticle at precisely the same rate,
so no baryon number can be generated.
\item An Arrow of Time (Departure from Thermal Equilibrium): The universe,
for much of its history, was very nearly in thermal equilibrium.
The spectrum of the CMBR is the most perfect blackbody spectrum
measured in nature.  So the universe was certainly in thermal
equilibrium $10^5$ years after the big bang.  The success of the
theory of big bang nucleosynthesis (BBN) provides strong evidence that the
universe was in equilibrium two-three minutes after the big
bang.  But if, through its early history, the
universe was in thermal equilibrium, then even $B$ and
$CP$ violating interactions could not produce a net asymmetry.
One way to understand this is to recall that the CPT theorem
assures strict equality of particle and antiparticle masses, so at
thermal equilibrium, the densities of particles
and antiparticles are equal.  More precisely,
since $B$ is odd under
CPT, its thermal average vanishes in an equilibrium situation.
This can be generalized by saying that the universe must have an
arrow of time.
\end{enumerate}

\smallskip
One of the great successes of the Standard Model is that it
explains why baryon and lepton number are conserved, to a very
good approximation.  To understand what this means,
consider first the modern understanding of
Maxwell's equations.  A quantum field theory is specified
by its field content and by a lagrangian density.
In the lagrangian, one distinguishes renormalizable and
non-renormalizable terms.  Renormalizable terms have coefficients
with mass dimension greater than zero; non-renormalizable terms
have coefficients (couplings) with mass dimension less than zero.
For example, in quantum electrodynamics, the electron mass has
dimension one, while the charge of the electron is dimensionless
(throughout we use conventions where $\hbar$ and $c$ are
dimensionless), so these
are renormalizable.  In fact, requiring Lorentz invariance, gauge invariance,
and renormalizability leaves only one possibility for the
lagrangian of electrodynamics:  the Maxwell lagrangian, whose
variation yields Maxwell's equations.   One can, consistent with
these
symmetry principles, write down an
infinite number of possible non-renormalizable terms, which would
yield non-linear modifications of Maxwell's equations.  There is
nothing wrong with these, but they are characterized by a mass, or
inverse length scale, $M$.   So the size of non-linear corrections
at wavelength $\lambda$ is of order $(\lambda M)^{-n}$ for some integer $n$.
$M$ represents
some scale at which the laws of electricity and magnetism might be
significantly modified.  Such corrections actually exist, and are
for most purposes quite small.

Similarly, in the Standard Model, at the level of renormalizable terms,
there are simply no interactions one can write which violate
either baryon number or the conservation of the separate
lepton numbers (electron, muon and tau number).
It is possible to add dimension five operators
(suppressed by some scale $1/M$) which violate lepton number, and
dimension six operators (suppressed by $1/M^2$), which violate
baryon number.  Again, these non-renormalizable terms must be
associated with some mass scale of some
new baryon and lepton violating
physics.  The dimension five lepton-number violating
operators would give rise to a mass for the neutrinos.
The recent
discovery of neutrino mass probably amounts to a measurement of some
of these lepton number violating operators. The scale of new physics
associated with these operators can not yet be determined, but
theoretical arguments suggest a range of possibilities, between
about $10^{11}$ and $10^{16}$ GeV.

The question, then, is what might be the scale, $M_{ B}$
associated with baryon number violation.
At the very least, one
expects quantum effects in gravity to violate
all global quantum numbers (e.g.
black holes swallow up any quantum numbers not connected with long
range fields like the photon and graviton), so $M_B \le M_p$, where
$M_p$, the Planck mass, is about $10^{19}$ GeV, or $(2 \times 10^{-33} {\rm
cm})^{-1}$.
The leading operators of this kind, if they have Planck mass
coefficients, would lead to a proton lifetime of order $10^{34}$
years or so.

If quantum gravitational effects were the only source of baryon number
violation, we could imagine that the baryon asymmetry of the universe was
produced when the temperature of the universe was of order the Planck
energy ($10^{32}\,^{o}$K).  Some complex processes associated with very
energetic configurations would violate baryon number.  These need not be in
thermal equilibrium (indeed, in a theory of gravity, the notion of
equilibrium at such a high temperature almost certainly does not make
sense).  The expansion of the universe at nearly the moment of the big bang
would provide an arrow of time.  CP is violated already at relatively low
energies in the Standard Model (through the Kobayashi-Maskawa
(KM) mechanism), so there is no reason to believe that it is conserved in
very high energy processes.  So we could answer Sakharov by saying that the
magnitude of the baryon number is the result of some very complicated,
extremely high energy process, to which we will never have experimental
access.  It might be, in effect, an initial condition.

There are good reasons to believe that this pessimistic
picture is not the correct one.
First, we are trying to understand a small, dimensionless number.
But in this Planck scale baryogenesis picture, it is not clear
how such a small dimensionless number might arise.
Second,  there is growing evidence that the universe
underwent a period of inflation early in its history.  During
this period, the universe expanded rapidly by an enormous factor
(at least $e^{60}$).   Inflation is likely to have taken
place well below the scale of quantum gravity, and thus
any baryon number produced in the Planck era was diluted to a totally
negligible level. Third, there are a variety of proposals for new physics --
as well as some experimental evidence -- which suggests that baryon
and lepton
number violating interactions might have been important at scales
well below the Planck scale.  So there is some reason for optimism
that we might be able to compute the observed baryon number
density from some underlying framework, for which we could provide
both direct (i.e. astrophysical or cosmological) and/or indirect
(discovery of new particles and interactions) evidence.

Several mechanisms have been proposed to understand the baryon
asymmetry:
\begin{enumerate}
\item  Planck scale baryogenesis:  this is the idea,
discussed above, that Planck scale phenomena are responsible for
the asymmetry.  We have already advanced arguments (essentially
cosmological) that this is unlikely; we will elaborate on them in
the next section.
\item  Baryogenesis in Grand Unified Theories (GUT baryogenesis):  this,
  the earliest 
well-motivated scenario for the origin of the asymmetry,
will be discussed more
thoroughly in the next section.  Grand Unified Theories unify the gauge
interactions of the strong, weak and electromagnetic
interactions in a single gauge group.  They inevitably violate
baryon number, and they have heavy particles, with
mass of order $M_{_{\rm GUT}}\approx 10^{16}$ GeV, whose decays
can provide a departure from equilibrium.  The main objections to this
possibility come from issues associated with inflation.  While there does
  not exist a 
compelling microphysical model for inflation, in most models,
the temperature of the universe after reheating is well
below $M_{_{\rm GUT}}$.  But even if it were very large, there
would be another problem.  Successful unification requires
supersymmetry, a hypothetical symmetry between fermions and bosons,
which will play an important role in this review.
Supersymmetry implies that the graviton
has a spin-3/2 partner, called the gravitino.   In most models for
supersymmetry breaking, these particles have masses of order
TeV, and are very long lived. Even though these particles are
weakly interacting, too many gravitinos are produced, unless
the reheating temperature is well below the unification
scale~\cite{kallosh}.
\item   Electroweak baryogenesis:  as we will explain,
the Standard Model satisfies all of the conditions for
baryogenesis.  This is somewhat surprising, since at low
temperatures the model seems to preserve baryon number, but it
turns out that baryon and lepton number are badly violated at very high
temperatures.  The departure from thermal equilibrium can arise
at the electroweak phase transition -- a transition between the
familiar state in which the $W$ and $Z$ bosons are massive and one
in which they are massless.  This transition can be first order, providing
an arrow of time.
It turns out, however, that as we will explain below,
any baryon asymmetry produced is {\it
far} too small to account for observations.  In certain extensions
of the Standard Model, it is possible to obtain an adequate
asymmetry, but in most cases the allowed region of parameter space
is very small.  This is true, for example, of the Minimal Supersymmetric
Standard Model (MSSM).  Experiments will soon either discover
supersymmetry in this region, or close off this tiny segment
of parameter space.

\item  Leptogenesis:  The observation that the weak interactions
will convert some lepton number to baryon number means that if one
produces a large lepton number at some stage, this will be
processed into a net baryon and lepton number.  The observation of
neutrino masses makes this idea highly plausible.  Many but not
all of the relevant parameters can be directly measured.
\item  Production by coherent motion of scalar fields (the
Affleck-Dine mechanism):  This mechanism, which can be highly
efficient, might well be operative if nature is supersymmetric.  In
this case, as we will explain in much greater detail,
the ordinary quarks and leptons are accompanied by
scalar quarks and leptons.  It has been widely conjectured
that supersymmetry may be discovered in the next generation
of high energy accelerators.  So again,
one might hope to uncover the basic underlying
physics, and measure some (but it will
turn out not all) of the relevant
parameters.  In non-supersymmetric theories,
it is believed that scalar fields with the requisite
properties (low mass, very flat potentials) are unnatural.
This supersymmetric baryogenesis mechanism will be the main focus of this
review. 
\end{enumerate}

In this review we will survey these, and explain in more detail
why the last two are by far the most plausible.
The question then becomes:  can we eventually establish that one or
the other
is correct?
In order to establish or rule out particular models for the origin
of the matter-antimatter asymmetry, we would hope to bring to bear
both astrophysical/cosmological and particle physics experiments and
observations, as well as theoretical arguments.
Ideally, we would some day be in the position of measuring
all of the parameters relevant to the asymmetry, and calculating
the asymmetry in much the same way that one presently calculates
the light element abundances.  One question we will ask is: how
close can we come to this ideal situation?


In the next section, after a very brief review of the standard cosmology,
we present our survey these mechanisms, both explaining how they work and
discussing their theoretical plausibility.  Both electroweak
baryogenesis and leptogenesis rely on the existence of processes within the
standard model which violate baryon and lepton number at high
temperatures, and we include a brief explanation of this
phenomenon.

We then turn to a more detailed discussion of
coherent production of baryons or leptons, the Affleck-Dine (AD)
mechanism.    This mechanism is potentially extremely efficient;
it can also operate relatively late in the history of the universe.
As a result, it can potentially resolve a number of cosmological
puzzles.  The AD mechanism presupposes low energy supersymmetry.
Supersymmetry (sometimes called SUSY for short)
is a hypothetical extension of Poincare invariance,
a symmetry which would relate bosons to fermions.  If correct, it
predicts that for every boson of the standard model, there is a
fermion, and vice versa.  It is believed that the masses of the
new particles should be about a TeV.  As supersymmetry
will play an important role in much of our discussion,
a brief introduction to
supersymmetry will be provided in the next section.
The supersymmetry hypothesis will be tested over the next decade
by the Tevatron and the Large Hadron Collider at CERN.
Interestingly, most other proposals for baryogenesis invoke
supersymmetry in some way (including electroweak baryogenesis and
most detailed models for leptogenesis).

\section{ A Baryogenesis Roadmap}

\label{sec:cosmooverview}
\subsection{A Cosmology Overview}

Our knowledge of the Big Bang rests on a few key observational elements.
First, there is the Hubble expansion of the universe.  This allows us to
follow the evolution of the universe to a few billion years after the Big
Bang.  Second, there is the CMBR.  This is a relic of the time, about
$10^5$ years after the Big Bang, when the temperature dropped to a fraction
of an electron volt and electrons and nuclei joined to form neutral atoms.
Third, there is the abundance of the light elements.  This is a relic of
the moment of neutrino decoupling, when the temperature was about $1$ MeV.
As we have noted, theory and observation are now in good agreement,
yielding the baryon to photon ratio of equation (\ref{baryondensity}).
Finally, there are the fluctuations in the temperature of the microwave
background, measured recently on angular scales below one degree by
BOOMERANG~\cite{CMBR_BOOMERANG_1,CMBR_BOOMERANG_2},
MAXIMA~\cite{CMBR_MAXIMA}, DASI~\cite{CMBR_DASI}, and WMAP~\cite{map}.
These fluctuations are probably a relic of the era of inflation (discussed
in more detail below).  The baryon density can be inferred independently
from the CMBR data and from the BBN determination of the baryon density
based on the measurements of the primordial deuterium
abundance~\cite{kirkman,burles}.  The agreement is spectacular: $\Omega_B
h^2 = 0.0214\pm 0.002$ based on BBN~\cite{kirkman}, while the CMBR
anisotropy measurements yield $\Omega_Bh^2 = 0.0224\pm 0.0009$~\cite{map}.

The first and perhaps most striking lesson of the measurements of the CMBR
is that the universe, on large scales, is extremely homogeneous and
isotropic.  As a result, it can be described by a Robertson-Walker metric:
\beq
ds^2 = dt^2 - R^2(t) \left ({dr^2 \over 1-kr^2} + r^2 d\theta^2 + r^2
\sin^2 \theta d\phi^2 \right ).\label{robertson}
\eeq
$R(t)$ is known as the scale factor; the Hubble ``constant" is
$H={\dot R \over R}$.  The Hubble constant is essentially the
inverse of the time; in the radiation dominated era, $H={1 \over
2t}$; in the matter dominated era, $H={2 \over 3 t}$.
It is puzzling that the universe should be homogeneous and
isotropic to such a high degree.  If one runs the clock
backward, one finds that vast regions of the universe which have
only recently been in causal contact have essentially the same
temperature.

Inflation provides an explanation for this and other
puzzles\cite{kolb,lindebook}.  The basic
idea~\cite{guth,linde_infl,steinhardt_infl}  is that for a
brief period, $R(t)$ grew extremely rapidly, typically exponentially.  This
has several effects:
\begin{itemize}
\item  The observed universe grew from a microscopically small
region, explaining homogeneity and isotropy.
\item  $k=0$, i.e. the universe is spatially flat.  This is now
well-verified by observations.
\item  Small fluctuations in the metric and the field during
inflation explain the observed small (part in $10^{-5}$) variation
in the temperature of the CMBR; detailed features of this
structure, in agreement with the inflationary theory, have
now been observed.  These fluctuations provide the seeds for
formation of the observed structure in the universe.
\item  Inflation also explains the absence from the universe of
objects such as magnetic monopoles expected in many particle
physics theories.
\end{itemize}

While it is probably fair to say that no
compelling microscopic theory of inflation yet exists,
as a phenomenological theory, inflation is very successful.
Most pictures of inflation invoke the dynamics of a scalar field
in a crucial way.  This scalar field must have very special
properties.  Typically, for example, the curvature of its
potential must be very small.
The most plausible theories which achieve this
invoke supersymmetry in a significant
way.  Supersymmetry, a hypothetical symmetry between fermions and
bosons, will be discussed at greater length later in this article.
It has been widely considered as a possible solution to many
puzzles in particle physics.  Most importantly for inflation,
supersymmetry is a theoretical framework which naturally gives
rise to scalars with very flat potentials.  It also, almost
automatically, gives rise to stable particles with just the right
properties to constitute the dark matter.  There are difficulties
as well.  One is associated with the fermionic partner of the
graviton, the gravitino.  In many models, this particle is very
long-lived ($\tau > 10^6$sec), and can spoil conventional Big Bang
cosmology if too many are produced.

There is not space here to review the subject of inflation.
Instead, we will give a ``narrative" of a possible history of the
universe, which will be useful to orient our discussion:
\begin{itemize}
\item  Before $t \approx 10^{-25}$ seconds, the universe was
very inhomogeneous, with an extremely large energy
density.  At $t \approx 10^{-25}$, inflation began in a
small patch.  This was associated with a scalar field, called the
inflaton, which moved slowly toward the minimum of its potential.
\item  The
scale of the inflaton potential was of order $10^{60}$ GeV$^4$, give
or take a few orders of magnitude.
\item
During inflation, the scale factor increased by an enormous
factor.  Any conserved or approximately conserved
 charges, such as monopole number or baryon number, were reduced
 by at least a factor of $10^{60}$ in this process
\item  Inflation ended as the inflaton approached the minimum of
its potential.  At this point, decays of the inflaton lead to
reheating of the universe to a high temperature.  Depending on the
detailed microscopic picture, there are constraints on the reheating
temperature.  If nature is supersymmetric, there is often a danger
of producing too many gravitinos and other long-lived particles.
Typically, this constrains the reheating temperature to be below
$10^{9}$ GeV.    Even without supersymmetry,
detailed inflationary models have difficulty producing high reheating
temperatures without fine tuning.
\item   The baryon asymmetry is generated some time after the era of
inflation.  Any upper limit on the reheating temperature
constrains the possible mechanisms for baryogenesis.
\end{itemize}

\label{sec:Grav}
\subsection{Planck Scale Baryogenesis}
\label{sec:GUTandL}

It is generally believed that a quantum theory of gravity cannot
preserve any global quantum numbers.  For example, in the collapse
of a star to form a black hole, the baryon number of the star is
lost; black holes are completely characterized by their mass,
charge and angular momentum.  Virtual processes involving black
holes, then, would also
be expected to violate baryon number.

In string theory, the only consistent quantum theory of gravity we
know, these prejudices are born out.  There are no conserved
global symmetries in string theory\cite{banksglobal}.
While we can't reliably extract detailed predictions from quantum
gravity for baryon number violation, we might expect that it will
be described at low energies by operators which appear in an
effective field theory.  The leading operators permitted by
the symmetries of  the Standard
model which violate baryon number carry dimension six.  An example
is:
\beq
{\cal L}_B = {1 \over M^2} \bar e \bar d^* \bar d^* \bar d^*
\label{bviolating}
\eeq
In this equation and those which follow, the various fermion fields, $d$
$\bar d$, $e$, $\bar e$, 
$\nu$, etc. are spinors of left-handed chirality.  $\bar d$ contains
the creation operator for the right-handed $d$ quark;
$d^\dagger$ for the left-handed anti-d quark.
The other two d-quark states
are created by $d$ and $d^{\dagger}$.
Because the operator is of dimension six, we have
indicated that its coefficient has dimensions of inverse
mass-squared.  This is analogous to the effective interaction in
the Fermi theory of weak interactions.  If quantum gravity is
responsible for this term, we might expect its coefficient to be
of order $1/M_p^2$, where $M_p = \sqrt{G_N^{-1}} = 10^{19} {\rm
GeV}$.

Because of this very tiny coefficient, these effects could be
important only at extremely early times in the universe, when, for
example, $H \sim M_p$.  It is probably very difficult to analyze
baryon production in this era.   It is certainly
unclear in such a picture where the small number $10^{-10}$
might come from.   But even if the baryon number was produced in
this era, it was completely washed out in the subsequent period of
inflation.   So gravitational baryogenesis seems unlikely to be
the source of the observed matter-antimatter asymmetry.

\subsection{GUT baryogenesis}
\label{subsec:GUT}

The earliest well-motivated scenarios for implementing Sakharov's ideas
within a detailed microscopic theory were provided by grand unified
theories (GUTs)\cite{kolb}.  In the Standard Model, the strong, weak and
electromagnetic interactions are described by non-abelian gauge theories
based on the groups $SU(3)$, $SU(2)$ and $U(1)$.  Grand unification
posits that the underlying theory is a gauge theory with a simple
group; this gauge symmetry is broken at some very high energy scale down to
the group of the Standard Model.  This hypothesis immediately provides an
explanation of the quantization of electric charge.  It predicts that, at
very high energies, the strong, weak and electromagnetic couplings
(suitably normalized) should have equal strength.  And most important, from
the point of view of this article, it predicts violation of baryon and
lepton numbers.

If nature is not supersymmetric, the GUT hypothesis fails.  One can use the
renormalization group to determine the values of the three gauge couplings
as a function of energy, starting with their measured values.  One finds
that they do not meet at a point, i.e. there is no scale where the couplings
are equal.  Alternatively, one can take the best
measured couplings, the $SU(2)$ and $U(1)$ couplings, and use the GUT
hypothesis to predict the value of the strong coupling.  The resulting
prediction is off by 12 standard deviations~\cite{pdg}.  But if one assumes
that nature is supersymmetric, and that the new particles predicted by
supersymmetry all have masses equal to 1 TeV, one obtains
unification, within 3 $\sigma$.  The scale of unification turns out to
be $M_{_{\rm GUT}}\approx 2\times 10^{16}$.  Relaxing the assumption that the
new particles are degenerate, or assuming that there are additional,
so-called threshold corrections to the couplings at the GUT
scale ($\sim4\%$) can yield complete agreement.

This value of $M_{_{\rm GUT}}$ is quite interesting.  It is sufficiently
below the Planck scale that one might hope to analyze these
theories without worrying about quantum gravity corrections.
Moreover, it leads to proton decay at a rate which may be
accessible to current proton decay experiments.  In fact, the
simplest SUSY GUT, based on the gauge group  $SU(5)$ is almost
completely
ruled out by the recent
Super-Kamiokande bounds~\cite{murayama}.  However, there are many
other models.  For example, non-minimal SU(5) or SO(10)
SUSY GUTs may have a proton lifetime about a factor of 5 above the present
experimental limit~\cite{raby,wilczek_gut,senjanovic,altarelli}.
Witten has recently advocated an approach to GUT model
building~\cite{witten_GUT,witten_GUT_2}
which resolves certain problems with these
models, and in which proton decay might be difficult to see even
in large detectors which are being considered for the future.

GUTs provide a framework which satisfies all three of Sakharov's conditions.
Baryon number violation is a hallmark
of these theories:  they typically contain gauge bosons and other
fields which mediate B violating interactions such as proton
decay.  CP violation is inevitable; necessarily, any
model contains at least the KM mechanism for violating CP,
and typically there are many new couplings
which can violate CP.  Departure from equilibrium is associated
with the dynamics of the massive, B-violating fields.  Typically
one assumes that these fields are in equilibrium at temperatures
well above the grand unification scale.  As the temperature
becomes comparable to their mass, the production rates of these
particles
fall below their rates of decay.  Careful calculations in these
models often leads to baryon densities compatible with what we
observe.

We can illustrate the basic ideas with the simplest GUT model, due
to~\textcite{gg}.
Here the unifying gauge group is $SU(5)$.  The model we will discuss
is not supersymmetric, but it illustrates the important
features of GUT baryon number production.  A single generation
of the standard model (e.g. electron, electron neutrino, u-quark, and d-quark)
can be embedded in the $\bar 5$ and $10$ representation
of $SU(5)$.  It is conventional, and convenient,
to treat all quarks and leptons as left-handed fields.  So in
a single generation of quarks and leptons one has the quark
doublet, $Q$, the singlet $\bar u$ and $\bar d$ antiquarks ({\it
their} antiparticles are the right-handed quarks),
and the lepton doublet, $L= \left (\matrix{e \cr \nu}\right ).$
Then
it is natural to identify the fields in the $\bar 5$
as
\begin{equation}
 \bar 5_i= \left ( \matrix{ \bar d \cr \bar d \cr \bar d \cr e \cr \nu}
\right ).
\label{fivebar}
\end{equation}
The generators of $SU(3)$ of color are identified as:
\beq
T =\left ( \matrix{ {\lambda^a \over 2} & 0 \cr 0 & 0} \right
)\label{suthreeembedding}
\eeq
where $\lambda^a$ are the Gell-Mann matrices,
while those of $SU(2)$ are identified with:
\beq
T =\left ( \matrix{ 0 & 0 \cr 0 & {\sigma^i \over
2}} \right )\label{sutwoembedding}
\eeq
The $U(1)$ generator is
\beq Y^\prime = {1 \over \sqrt{60}}\left (\matrix{2 & & & & \cr & 2 & & &
\cr & & 2 && \cr & & & -3 & \cr & & & & -3} \right
).\label{hyperchargeprime}
\eeq
Here the coefficient has been chosen so
that the normalization is the same as that of the $SU(3)$ and $SU(2)$
matrices (${\rm Tr}(T^a T^b)=\delta_{ab}$).  The corresponding
gauge boson couples with the same coupling constant as the gluons and $W$
and $Z$ bosons.  This statement holds at $M_{_{\rm GUT}}$; at lower energies,
there are significant radiative corrections (which in the
supersymmetric case reproduce the observed low energy gauge
couplings).

In the standard model, the hypercharge is related to the ordinary electric
charge, $Q$, and the isospin generator, $T_3$, by $Q=T_3 + {Y \over 2}$.
So one sees that electric charge
is quantized, and that $Y= \sqrt{3/40}~Y^\prime$.  Since $Y$ couples with
the same
strength as the $SU(2)$ generators, this gives a prediction of the $U(1)$
coupling of the standard model, and correspondingly of the Weinberg angle,
$\sin^2(\theta_W) = 3/8$.  This prediction receives radiative corrections,
which, assuming supersymmetry, bring it within experimental errors of the
measured value.

In a single generation, the remaining
fields lie in the $10$ representation.  The $10$ transforms as the
antisymmetric product of two $5$'s.  It has the form
\beq
10_{ij} = \left ( \matrix{0 & \bar u_2 & -\bar u_1 & Q_1^1 & Q_1^2 \cr
-\bar u_2 & 0 & \bar u_3 & Q_2^1 & Q_2^2 \cr \bar u_1 & -\bar u_3 & 0 & Q_3^1 &
Q_3^2 \cr -Q_1^1 & -Q_2^1 & -Q_3^1 & 0 & \bar e \cr -Q_1^2 & -Q_2^2 &
-Q_3^3 & -\bar e & 0 } \right ),\label{tenrepresentation}
\eeq
where $Q^{i}=(u^i,d^i)$ are left-handed quark fields, which transform as
doublets under the SU(2).  $SU(5)$ is not a manifest symmetry of nature.
It can be broken by the expectation value of a scalar field in the adjoint
representation with the same form as $Y$:
\beq
\langle \Phi \rangle 
= v \left (\matrix{2 & & & & \cr & 2 & & & \cr & & 2
&&
\cr & & & -3 & \cr & & & & -3} \right ).\label{adjointvev}
\eeq
The unbroken generators are those which commute with $\Phi$, i.e.
precisely the generators of $SU(3)\times SU(2) \times U(1)$ above.

The vector bosons which correspond to the broken generators gain mass of
order $g v$.  We will refer to the corresponding gauge bosons as $X$; they
are associated with generators which don't commute with $<\Phi>$, such as:
\beq
\left (\matrix{ 0 & 0 & 0 & 1 & 0 \cr 0 & 0 & 0& 0 & 0 \cr 0 & 0 & 0 & 0 &
  0 \cr
1& 0 & 0 & 0  & 0 \cr 0 & 0 & 0 & 0 & 0 } \right ).
\eeq
They carry color and electroweak quantum numbers and mediate processes which
violate baryon number.  For example, examining the structure of
\ref{fivebar}, one sees that there is a coupling of the $X$ bosons to a
$\bar d$ quark and an electron.  Similarly, there is a coupling of the $X$
boson to a quark doublet and a positron.  Note that there is no way to
assign baryon and lepton number to the $X$ boson so that it is conserved by
these couplings.

In the GUT picture of baryogenesis, it is usually assumed that at
temperatures well above the GUT scale, the universe was in thermal
equilibrium.  As the temperature drops below the mass of the $X$ bosons,
the reactions which produce the $X$ bosons are not sufficiently rapid to
maintain equilibrium.  The decays of the $X$ bosons violate baryon number;
they also violate CP.  So all three conditions are readily met: baryon
number violation, CP violation, and departure from equilibrium.

To understand in a bit more detail how the asymmetry can come about, note that
CPT requires that the total decay rate of $X$ is the same as that of its
antiparticle  $\bar
X$.  But it does not require equality of the decays to particular final
states (partial widths).  So starting with equal numbers of $X$ and $\bar
X$ particles, there can be a slight asymmetry between the processes
\beq
X \rightarrow d L; X \rightarrow \bar Q \bar u
\label{xdecayone}
\eeq
and
\beq
\bar X \rightarrow \bar d \bar L; \bar X \rightarrow Q u.\label{xdecaytwo}
\eeq
The tree graphs for these
processes are necessarily equal; any CP violating phase simply
cancels out when we take the absolute square of the amplitude.
This is not true in higher order, where additional phases
associated with real intermediate states can appear.
Actually computing
the baryon asymmetry requires a detailed analysis, of a kind we
will encounter later when we consider leptogenesis.

\begin{figure}
\centerline{\epsfxsize=5.5in \epsfbox{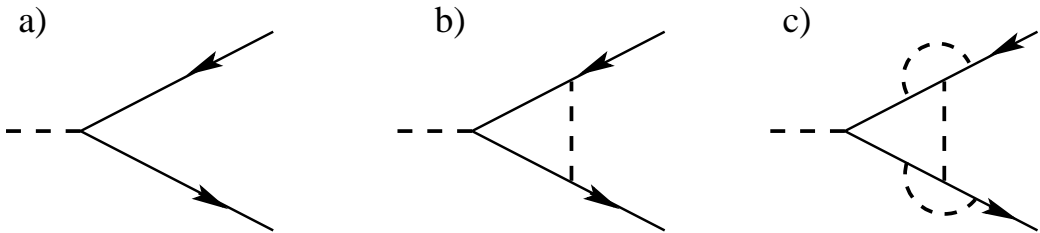}}
\caption{ Interference between the tree-level (a) and one-loop (b) diagrams
with complex Yukawa couplings can provide the requisite source of CP
violation for GUT baryogenesis.  In viable models, to avoid the unwanted
cancellations, one must often assume that the two scalars are different or
go to higher loops (c)~\cite{barr,kolb}.  }
\label{feynman}
\end{figure}

There are reasons to believe, however, that GUT baryogenesis is not the
origin of the observed baryon asymmetry.  Perhaps the most compelling of
these has to do with inflation.  Assuming that there was a period of
inflation, any pre-existing baryon number was greatly diluted.  So in order
that one produce baryons through $X$ boson decay, it is necessary that the
reheating temperature after inflation be at least comparable to the $X$
boson mass.  But as we have explained, a reheating temperature greater than
$10^9$ GeV leads to cosmological difficulties, especially overproduction of
gravitinos.

\subsection{Electroweak Baryon Number Violation}
\label{sec:EBNV}

Earlier, we stated that the renormalizable interactions of the Standard
Model preserve baryon number.  This statement is valid classically, but it
is not quite true of the quantum theory.  There are, as we will see in this
section, very tiny effects which violate baryon number~\cite{thooft}.
These effects are
tiny because they are due to quantum mechanical tunneling, and are
suppressed by a barrier penetration factor.  At high temperatures, there is
no such suppression, so baryon number violation is a rapid process, which
can come to thermal equilibrium.  This has at least two possible
implications.  First, it is conceivable that these ``sphaleron" processes
can themselves be responsible for generating a baryon asymmetry.  This is
called electroweak baryogenesis~\cite{krs}.  Second, as we will see, sphaleron
processes can process an existing lepton number, producing a net lepton and
baryon number.  This is the process called leptogenesis~\cite{fy}.

In this section, we summarize the main arguments that the
electroweak interactions violate baryon number at high temperature.
In the next section, we explain why the electroweak interactions might
produce a small
baryon excess, and why this excess cannot be large enough to
account for the observed asymmetry.

One of the great successes of the Standard
Model is that it
explains the observed conservation laws.  In particular, there are
no operators of dimension four or less consistent with
the gauge symmetries which violate baryon number
or the separate lepton numbers.  The leading operators which can
violate can baryon number are of dimension six, and thus suppressed by
${\cal O}({1 \over M^2})$.  The leading operators which violate the
separate lepton numbers are of dimension five, and thus suppressed
by one power of $1/M$.  In each case, $M$ should be thought of as
the scale associated with some very high energy
physics which violates baryon or lepton number.
It cannot be determined except through measurement or
by specifying a more microscopic theory.

However, it is not quite true that the standard model preserves
all of these symmetries.  There are tiny effects, of order
\beq
e^{-(2 \pi / \alpha_{\rm _W}) }\approx 10^{-65}
\eeq
which violate them.  These effects are related to the fact
that the separate baryon number and lepton number currents are
``anomalous."   When one quantizes the theory carefully, one finds
that the baryon number current, $j^{\mu}_B$,  is not exactly conserved
but rather satisfies:
\beq
\partial_{\mu}j^{\mu}_B = {3 \over 16 \pi^2} F_{\mu \nu}^a
\tilde F_{\mu \nu}^a = {3 \over 8 \pi^2} {\rm Tr} F_{\mu \nu}
\tilde F_{\mu \nu}. \label{axialanomaly}
\eeq
Here $F_{\mu \nu}$ are the $SU(2)$ field strengths, and
we have introduced matrix-valued fields in the last
expression,
\beq
F_{\mu \nu}=\sum_a F_{\mu \nu}^a T^a,
\eeq
and similarly
for other fields, and the dual of $F$, $\tilde F$, is defined by:
\beq
\tilde F_{\mu \nu} = {1 \over 2} \epsilon_{\mu\nu\rho\sigma}F^{\rho\sigma} .
\label{fdual}
\eeq
In electromagnetism, $F \tilde F = 2 \vec E \cdot \vec B$.

The same anomaly (\ref{axialanomaly}) appears in the lepton number current
as well, i.e.,
\beq
\partial_{\mu}j^{\mu}_L = {3 \over 16 \pi^2} F_{\mu \nu}^a
\tilde F_{\mu \nu}^a = {3 \over 8 \pi^2} {\rm Tr} F_{\mu \nu}
\tilde F_{\mu \nu}. \label{axialanomaly_L}
\eeq
However, the difference of the two, $j^{\mu}_B-j^\mu_L$, is anomaly-free and is
an exactly conserved quantity in the Standard Model (as well as SU(5) and
SO(10) grand unified theories).

One might think that such a violation of current conservation
would lead to dramatic violations of the symmetry.  But the
problem is more subtle.  The right hand side of the anomaly
equation is itself a total divergence:
\beq
{\rm Tr} F_{\mu \nu} \tilde F_{\mu \nu} = \partial_{\mu}
K^{\mu}\label{fftilde}
\eeq
where
\beq
K^{\mu} = \epsilon^{\mu \nu \rho \sigma} tr[F_{\nu \rho} A_{\sigma}
+{2 \over 3} A_{\nu} A_{\rho} A_{\sigma}]\label{kequals}
\eeq
(the reader can quickly check this for a $U(1)$ gauge theory like
electromagnetism).
In view of this,
\beq
\tilde j = j_B^{\mu} - {3 g^2 \over 8 \pi^2} K^{\mu} \label{jtilde}
\eeq
is conserved.
In perturbation theory (i.e. in Feynman diagrams), $K^{\mu}$
falls to zero rapidly (typically like $1/r^6$) at $\infty$, and so
its integral is zero. This fact insures that baryon number is
conserved.

In abelian gauge theories, this is the end of the story.  In
non-abelian theories, however, there are non-perturbative
field configurations
which contribute to the right hand side.  These lead to violations
of baryon number and the separate lepton numbers proportional to
$e^{-{2 \pi \over \alpha}}$.  These configurations are called
instantons.  We will not discuss them in detail here; a pedagogical
treatment is given by~\textcite{colemanuses}.  They
correspond to calculation of a tunneling amplitude.  To understand
what the tunneling process is, one must consider more carefully
the ground state of the field theory.  Classically, the ground
states are field configurations for which
the energy vanishes.  The trivial solution of
this condition is $\vec A = 0$, where $\vec A$ is the
vector potential.  More generally, one can
consider $\vec A$ which is a ``pure gauge,"
\beq
\vec A = {1 \over i} g^{-1} \vec \nabla g,
\eeq
where $g$ is a gauge transformation matrix.  In an abelian (U(1)) gauge
theory, fixing the gauge eliminates all but the trivial solution, $\vec
A=0$.\footnote{More precisely, this is true in axial gauge.  In the gauge
$A_o=0$, it is necessary to sum over all time-independent transformations
to construct a state which obeys Gauss's law.}  This is not the case for
non-abelian gauge theories.  There is a class of gauge transformations,
labeled by a discrete index $n$, which do not tend to unity as $\vert \vec
x \vert \rightarrow \infty$, which must be considered to be distinct
states.  These have the
form:
\beq
g_n(\vec x)= e^{i n f(\vec x) \hat x \cdot \tau/2}\label{large}
\eeq
where $f(x) \rightarrow 2 \pi$ as $\vec x \rightarrow \infty$,
and $f(\vec x) \rightarrow 0$ as $\vec x \rightarrow 0$.

So the ground states of the gauge theory are labeled by
an integer $n$.  Now if we evaluate the integral of the
current $K^o$, we obtain a quantity known as the Chern-Simons
number:
\beq
n_{_{CS}} = {1 \over 16 \pi^2} \int d^3 x K^o =
{2/3 \over 16 \pi^2} \int d^3 x \epsilon_{ijk} Tr (g^{-1} \partial_i g
g^{-1} \partial_j g g^{-1} \partial_k g).\label{csn}
\eeq
For $g=g_n$, $n_{_{CS}}=n$.  The reader can also check that for
$g^{\prime} = g_n(x)h(x)$, where $h$ is a gauge transformation which
tends to unity at infinity (a so-called ``small gauge transformation"),
this quantity is unchanged.  $n_{_{CS}}$, the ``Chern-Simons number,"
is topological in this sense (for $\vec A$'s which are not ``pure
gauge," $n_{_{CS}}$ is in no sense quantized).

\begin{figure}
\centerline{\epsfysize=2in \epsfxsize=3in \epsfbox{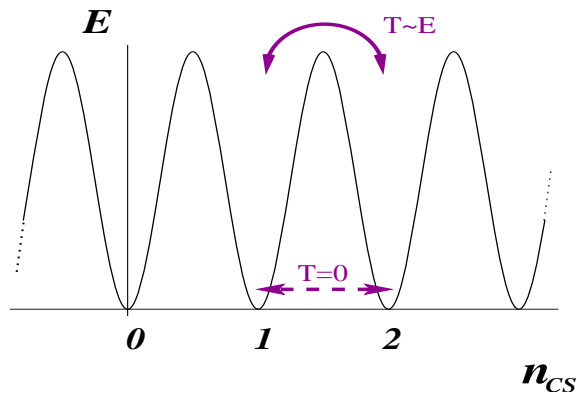}}
\caption{
Schematic Yang-Mills vacuum structure.  At zero temperature, the instanton
transitions between vacua with different  Chern-Simons numbers are
suppressed.  At finite temperature, these transitions can proceed via
sphalerons.
}
\label{Ncs}
\end{figure}

Schematically, we can thus think of the vacuum structure of a
Yang-Mills theory as indicated in Fig. \ref{Ncs}.  We have, at weak
coupling, an infinite set of states, labeled by integers, and
separated by barriers from one another.  In tunneling processes
which change the Chern-Simons number, because of the anomaly, the
baryon and lepton numbers will change.  The exponential
suppression found in the instanton calculation is typical
of tunneling processes, and in
fact the instanton calculation which leads to the result for the amplitude
is nothing but a field-theoretic WKB calculation.

At zero temperature, the decay amplitude is suppressed, not only by $e^{-2 \pi
\over \alpha}$, but by factors of Yukawa couplings.    The probability
that a single proton has decayed through this process in the
history of the universe is infinitesimal.  But this picture
suggests that, at finite temperature, the rate should be larger.
One can determine the height of the barrier separating configurations of
different $n_{_{CS}}$ by looking for the field configuration
which corresponds to sitting on top of the
barrier.  This is a solution
of the static equations of motion with finite energy.
It is known as a ``sphaleron"\cite{manton}.
When one studies the small fluctuations about this solution, one
finds that there is a single negative mode, corresponding to the
possibility of rolling down hill into one or the other well.  The
sphaleron energy is of order
\beq
E_{sp} = {c \over g^2}M_{_W}.
\eeq
This can be seen by scaling arguments on the
classical equations; determining the coefficient
$c$ requires a more detailed
analysis.  The rate for thermal fluctuations to cross
the barrier per unit time per unit
volume should be of order the
Boltzmann factor for this configuration,
times a suitable prefactor\cite{krs,arnoldmclerran,dinetal},
\beq
\Gamma_{sp}= T^4  e^{-E_{sp}/T}.\label{lowtrate}
\eeq
Note that the rate becomes large as the temperature approaches the $W$
boson mass.  In fact, at some temperature the weak interactions
undergo a phase transition to a phase in which the $W$ boson mass
vanishes.  At this point, the computation of the transition rate
is a difficult problem -- there is no small parameter -- but
general scaling arguments show that the transition rate is of the
form\footnote{more detailed considerations alter slightly the parametric
form of the rate~\cite{asy}}:
\beq
\Gamma_{\rm bv}= \alpha_{_W}^4 T^4.\label{hightrate}
\eeq

Returning to our original expression for the anomaly, we see that
while the separate baryon and lepton numbers are violated in
these processes, the combination $B-L$ is conserved.  This result
leads to three observations:

\begin{enumerate}
\item  If in the early universe, one creates baryon and lepton
number, but no net $B-L$, $B$ and $L$  will subsequently be lost
through
sphaleron processes.
\item  If one creates a net $B-L$ (e.g. creates a lepton
number) the sphaleron process
will leave both baryon and lepton numbers comparable to the
original $B-L$.  This realization is crucial to the idea of
leptogenesis, to be discussed in more detail below.
\item  The standard model satisfies, by itself, all of the
conditions for baryogenesis.
\end{enumerate}

\subsection{Electroweak baryogenesis}
\label{sec:EW}

As we will see, while the Standard Model satisfies all of the conditions
for baryogenesis~\cite{krs}, nothing like the required baryon number can be
produced.  It is natural to ask whether extensions of the Standard Model,
such as theories with complicated Higgs, or the Minimal Supersymmetric
(extension of the) Standard Model, can generate an asymmetry, using the
sphaleron process discussed in the previous section.  We will refer to such
a possibility more generally as ``Electroweak Baryogenesis."

\subsubsection{Electroweak baryogenesis in the Standard Model}
\label{sec:EWSM}

How might baryons be produced in the Standard Model?  From our
discussion, it is clear that the first and second of Sakharov's
conditions are satisfied.  What about the need for a departure
from equilibrium?

Above we alluded to the fact that in the electroweak theory, there
is a phase transition to a phase with massless gauge bosons.  It
turns out that, for a sufficiently light Higgs, this transition is
first order.
At zero temperature, in the simplest version of the Standard
Model with a single Higgs field, $\phi$, the Higgs potential is given by
\beq
V(\Phi) = -\mu^2 \vert \Phi \vert^2 + {\lambda \over 2}
\vert \Phi \vert^4.\label{zerotemp}
\eeq
The potential has a minimum at $\Phi= {1 \over \sqrt 2} v_o$,
breaking the gauge symmetry
and giving mass to the gauge bosons by the Higgs mechanism.

What about
finite temperatures?  By analogy with the phase transition
in the Landau-Ginsburg model of superconductivity,
one might expect that the value of $<\Phi>$ will change as the
temperature increases.  To determine the value of $\Phi$,
one must compute the free
energy as a function of $\Phi$.The leading
temperature-dependent corrections are obtained by simply
noting that the masses of the various fields in the theory --
the $W$ and $Z$ bosons and the Higgs field, in particular, depend
on $\Phi$.  So the contributions of each species to the
free energy are $\Phi$-dependent:
\beq
{\cal F}(\Phi)V_T(\Phi) = \pm \sum_i \int{ d^3 p \over 2 \pi^3} \ln \left (
1 \mp e^{-\beta 
\sqrt{p^2 + m_i^2(\Phi)}} \right )\label{vt}
\eeq
where $\beta = 1/T$, $T$ is the temperature, the sum is over all particle
species (physical helicity states), and the plus sign is for bosons, the
minus for fermions. In the Standard Model, for temperature $T\sim
10^2\,$GeV, one can treat all the quarks are massless, except for the
top quark.  The effective potential (\ref{vt}) then depends on the top
quark mass, $m_{_t}$, the vector boson masses, $M_{_Z}$ and $m_{_W}$, and
on the Higgs mass, $m_{_H}$.  Performing the integral in the equation
yields
\beq
V(\Phi,T) = D (T^2 - T_o^2) \Phi^2 - E T \Phi^3 +
{\lambda\over 4} \Phi^4 \ +\dots .
\label{hightemp}
\eeq
The parameters $T_o$, $D$ and $E$ are given in terms of the gauge
boson masses and the gauge couplings below.  For the moment,
though, it is useful to note certain features of this expression.
$E$ turns out to be a rather small, dimensionless number, of order
$10^{-2}$.  If we ignore the $\phi^3$ term, we have a second order
transition, at temperature $T_o$, between a phase with $\phi \ne
0$ and a phase with $\phi=0$.  Because the $W$ and $Z$ masses are
proportional to $\phi$, this is a transition between a state with
massive and massless gauge bosons.

\begin{figure}
\centerline{\epsfysize=2.5in \epsfxsize=4.5in \epsfbox{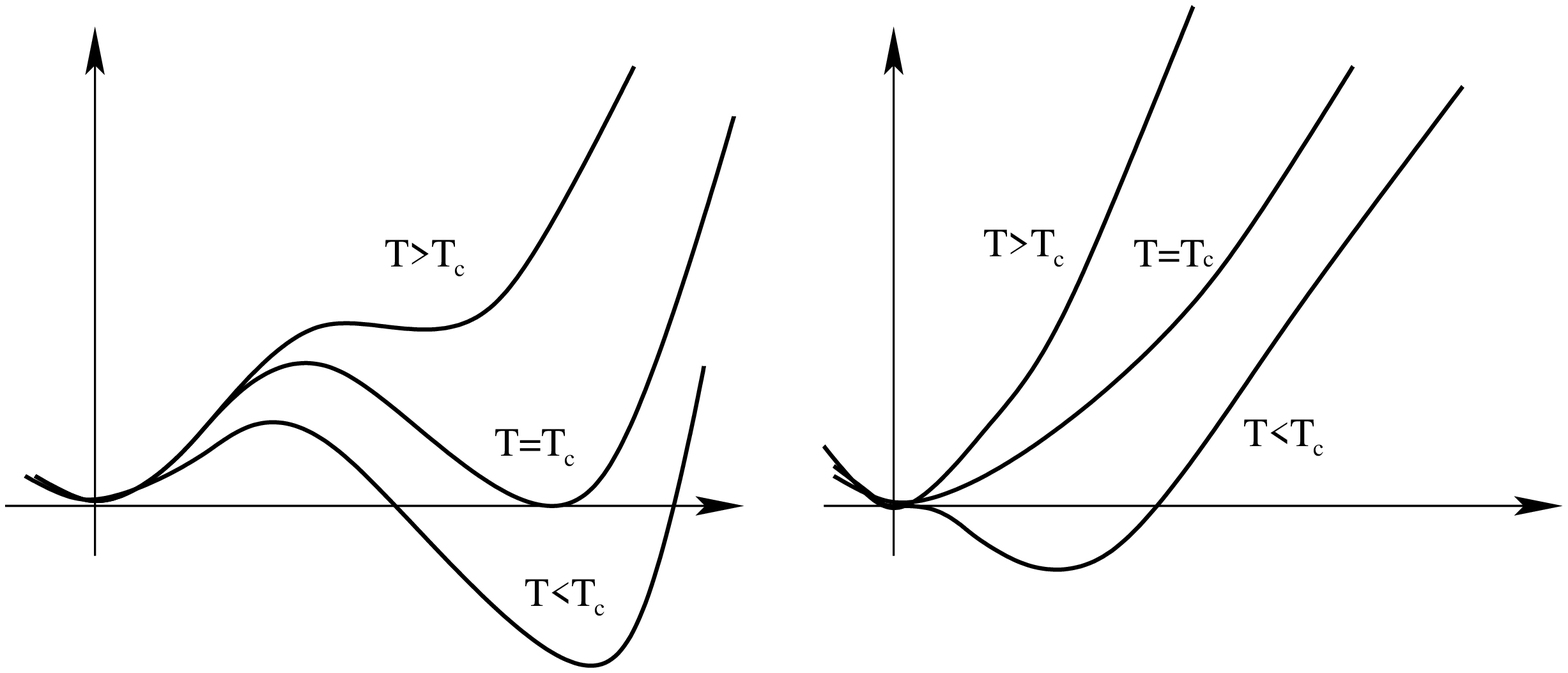}}
\caption{First and second order phase transitions.
}
\label{PT}
\end{figure}

Because of the $\phi^3$ term in the
potential, the phase transition is potentially at least weakly first order.
This is
indicated in Fig.~\ref{PT}.  Here one sees the appearance of
a second, distinct, minimum at a critical
temperature.  A first order transition is not, in general, an
adiabatic process.  As we lower the temperature to the transition
temperature, the transition proceeds by the formation of bubbles; inside
the bubble the system is in the true equilibrium state (the state
which minimizes the free energy) while outside it tends
to the original state.  These bubbles form through thermal flucutations
at different points in the system, and grown until they collide,
completing the phase transition.  The
moving bubble walls are regions where the Higgs fields are changing,
and
all of Sakharov's conditions are satisfied.  It has been shown
that various non-equilibrium processes near the wall can
produce baryon and lepton numbers~\cite{rs_review,ckn_review}.

Describing these processes would take us far afield.  Even without
going through
these details, however, one
point is crucial:
after the bubble has passed any given region, the baryon
violating processes should shut off.  If these processes continue, they
wash out the baryon asymmetry produced during the phase transition.

Avoiding the washing out of the asymmetry requires that after the
phase
transition, the sphaleron rate should be small compared to the
expansion rate of the universe.  According to
eq.~(\ref{lowtrate}), this
requires that after the
transition the sphaleron energy, which is proportional to
(temperature-dependent) $W$-boson mass, $M_{_W}$, be large compared
to the temperature.  This, in turn, means that the Higgs expectation value
must be 
large immediately after the transition.  Using eq.~(\ref{hightemp}) or more
refined calculations to higher orders, one can relate the change in
the Higgs expectation value to the Higgs mass at zero temperature.  It
turns out that the current lower limit on the Higgs boson mass rules out
any
possibility of a large enough Higgs expectation value immediately
after the
phase transition, at least in the minimal model with a single Higgs
doublet.

The shape of the effective potential $V(\phi,T)$ near the critical
temperature $T_c$ determines whether the phase transition is first order,
which is a necessary condition for electroweak baryogenesis to work.
Eq.~(\ref{hightemp}) represents the lowest order term in perturbation
theory; higher order terms have been computed as
well~\cite{twoloop_1,twoloop_2,twoloop_3,twoloop_4}.  However, these
calculations are not reliable, because of infrared divergences
which arise in the perturbation expansion.  These arise because the
Higgs field is nearly massless at the transition.
Numerical simulations are required, often
combined with a clever use
of perturbation theory~\cite{loops_and_lattice}.
Numerical simulations~\cite{EW_num1,EW_num2,EW_num3,EW_num_B,EW_l1,EW_l2,
EW_l3} have shown that, for the Higgs mass above 80~GeV (which it must be,
to satisfy the present experimental constraints), the sharp phase
transition associated with the low mass Higgs
turns into a smooth crossover.

However, even for an unrealistically light Higgs, the actual production of
baryon asymmetry in the minimal Standard Model would be highly suppressed.
This is because Standard Model CP violation must involve all three
generations~\cite{ckm}.  The lowest order diagram that involves three
generations of fermions with proper chiralities and contributes to CP
violating processes relevant to baryogenesis is suppressed by 12 Yukawa
couplings~\cite{ms86,ms87}.  Hence, the CKM CP violation contributes a
factor of $10^{-20}$ to the amount of baryon asymmetry that could arise in
the Standard Model.

Clearly, one must look beyond the Standard Model for the origin of baryon
asymmetry of the universe.  One of the best motivated candidates for new
physics is supersymmetry.

Before closing this section,
for completeness, we give the values of the parameters $T_o$, $B$,
$D$ and $E$ in eqn.~\ref{hightemp}:
\beq
T^2_o = {1\over 2D}(\mu^2 - 4Bv_o^2) =
{1\over 4D}(m_H^2 - 8Bv_o^2) \ ,
\label{tzero}
\eeq
while the parameters $B$, $D$ and $E$ are given by:
\beq
B= {3 \over 64 \pi^2 v_o^4}(2M_W^4 + M_Z^4 -4m_t^4)\, , ~~~~
D = {1\over 8v_o^2} ( 2 M_W^2 + m_Z^2 + 2 m_t^2) \, ,~~~~~
E =  {1\over 4\pi v_o^3} ( 2 M_W^3 + m_Z^3) \sim 10^{-2} .
\label{deequals}
\eeq

\subsubsection{Supersymmetry,  a short introduction}
\label{sec:SUSY}

In this section we provide
a brief introduction to supersymmetry.  Much more detail
can be found, e.g., in \cite{brush}
and in many texts.

There are many hints
that supersymmetry, a hypothetical symmetry
between fermions and bosons, might play some role in nature.  For
example, supersymmetry seems to be an essential part
of superstring theory, the only consistent theory of quantum
gravity which we know.  If supersymmetry is a symmetry
of the laws of nature, however, it must be badly broken; otherwise
we would have seen, for example, scalar electrons (``selectrons")
and fermionic photons (``photinos").  It has been widely conjectured
that supersymmetry might be discovered by accelerators
capable of exploring the TeV energy range.  There are several
reasons for this.  The most compelling is the ``hierarchy problem."
This is, at its most simple level, the puzzle of the wide disparity of
energies between the Planck scale (or perhaps the unification scale)
and the weak scale -- roughly $17$ orders of magnitude.
While one might take this as simply a puzzling fact, within quantum
theory, the question is made sharper by the fact that scalar masses
(particularly
the Higgs mass) receive very divergent quantum corrections.  A typical
expression for the quantum corrections to a scalar mass is:
\begin{equation}
\delta m^2 = {\alpha \over \pi} \int d^4 k{1 \over k^2}.
\label{deltam2}
\end{equation}
This integral diverges quadratically for large momentum ($k$).  Presumably,
the integral is cut off by some unknown physics.  If the energy scale of
this physics is $\Lambda$, then the corrections to the Higgs mass are {\it
much} larger than the scale of weak interactions unless $\Lambda \sim {\rm
TeV}$.  While various cutoffs have been proposed, one of the most
compelling suggestions is that the cutoff is the scale of supersymmetry
breaking.  In this case, the scale must be about $1000$ GeV.  If this
hypothesis is correct, the Large Hadron Collider under construction at CERN
should discover an array of new particles and interactions (it is possible
that supersymmetry could be discovered at the Tevatron beforehand).

The supersymmetry generators, $Q_{\alpha}$, are fermionic operators.
Acting on bosons they produce fermions degenerate in energy;
similarly, acting on fermions, they produce degenerate bosons.
Their algebra involves the total energy and momentum,
\beq
\{Q_{\alpha},Q_{\beta}\} = P^\mu \gamma^\mu_{\alpha \beta}.
\eeq
Neglecting gravity, supersymmetry is a global symmetry.  Because of the
structure of the algebra, the symmetry is
broken if and only if the energy of the ground state is non-zero.
If the symmetry is unbroken, for every boson there is a degenerate
fermion, and conversely.

If we neglect gravity, there are two types of supermultiplets which
may describe light fields.
These are the chiral multiplets,
containing a complex scalar and a Weyl (two-component) fermion
\beq
\Phi_i = (\phi_i, \psi_i),\label{chiralmultiplet}
\eeq
and the
vector multiplets, containing a gauge boson and a Weyl
fermion:
\beq
V^a =  (A_{\mu}^a, \lambda^a).\label{vectormultiplet}
\eeq
In global supersymmetry, the lagrangian is specified
by the gauge symmetry and an analytic (more precisely
holomorphic) function of the scalar fields,
$W(\phi_i)$,
known as the superpotential.  For renormalizable
theories,
\beq
W(\phi_i)= {1\over 2} m_{ij} \phi_i \phi_j
+ \lambda_{ijk} \phi_i \phi_j \phi_k.
\label{wrenorm}
\eeq
The lagrangian then includes the following:
\begin{enumerate}
\item  The usual covariant kinetic terms for all of the
fields, for example
\beq
\lambda^a D \lambda^{a*},~~~\psi D \psi^*,~~~
{-1\over 4}F_{\mu\nu}^{a~2},~~~\vert D_{\mu} \phi \vert^2.
\label{kinetic}
\eeq

\item  Yukawa couplings with gauge strength:
\beq
\sqrt{2}g^a \lambda^a \phi^*  T^a \psi + c.c.\label{yukawacouplings}
\eeq

\item  Mass terms and Yukawa couplings from $W$:
\beq
{1\over 2}{\partial^2 W \over \partial \phi_i \partial \phi_j}
\psi_i \psi_j
\label{componenty}
\eeq
\beq
~~~~~~~=m_{ij} \psi_i \psi_j + {3 \over 2} \lambda_{ijk} \phi_i
\psi_j \psi_k.
\label{componentl}
\eeq

\item  A scalar potential:
\beq
V= \sum_i \left \vert {\partial W \over \partial \phi_i} \right
\vert^2 +\sum_a
{1 \over 2} g^{a2} \left (\sum_i \phi_i^* T^a \phi_i \right )^2
\label{potential}
\eeq
It is convenient to define two types of auxiliary fields, the
``F" and ``D" fields:
\beq
  F_i={\partial W \over \partial \phi_i}~~~~~ D^a=g^a
\sum_i \phi_i^* T^a \phi_i.
\label{auxiliary}
\eeq
\end{enumerate}

In terms of these, the potential is simply
\beq
V=\vert F_i \vert^2 +{1\over 2} \vert D^{a} \vert ^{2}
\label{potedf}
\eeq
and, at the classical level, supersymmetry is unbroken
if and only if all of the $D$ and $F$ fields vanish at
the minimum of the potential.

\smallskip

It is useful to consider some examples.  Take first a
model with a single chiral field, $\phi$, and superpotential
\beq
W={1 \over 2}m \phi^2.
\label{simplemodel}
\eeq
In this case, the potential is
\beq
V= \left \vert {\partial W \over \partial \phi} \right \vert^2
= m^2 \vert \phi \vert^2.
\label{simplepot}
\eeq
On the other hand, the fermion mass is
just $m$, from eqn. (\ref{componentl}), so the model describes
two bosonic and two fermionic degrees of freedom, degenerate
in mass.

A more interesting model is a supersymmetric version
of the standard model, known as the Minimal Supersymmetric
Standard Model (MSSM).  The gauge group is
$SU(3) \times SU(2) \times U(1)$, and there is one
vector multiplet for each gauge generator.  In addition,
for each of the usual fermions of the standard model,
one has a chiral field with the same quantum numbers:
\beq
Q_a =(3,2)_{1/3} ~~~~\bar u_a = (\bar 3,2)_{-4/3}
~~~~~\bar d_a =  (\bar 3,2)_{2/3}~~~~~L_a=(1,2)_{-1}
~~~~~\bar e=(1,1)_2.
\label{quantumnos}
\eeq
Here $a$ is a generation index, $a=1,\dots,3$.
In addition, there are two Higgs fields (two in order to
cancel anomalies and to be able to give mass to all
quarks and leptons)
\beq
H_u=(1,2)_1~~~~~H_d=(1,2)_{-1}\label{higgsnos}
\eeq
In this notation, $(\chi,\psi)_{_Y}$ shows the transformation properties of
the fields under the action of the gauge group: $\chi$ and $\psi$ refer to
the representations of the SU(3) and SU(2) respectively, while $Y$ denotes
the hypercharge.

The superpotential of the model is a generalization of the
Yukawa couplings of the standard model:
\beq
W=\Gamma^{ab} Q_a \bar u_b H_u+ \gamma^{a} Q_a
\bar d_a H_d + \Gamma_a L_a \bar e_a H_d
+ \mu H_u H_d,\label{mssmw}
\eeq
where  we have used
our freedom to make field redefinitions to make
the (somewhat unconventional, but
later convenient) choice that the
$d$-quark and the lepton Yukawa couplings
are diagonal.  If one supposes that the Higgs fields $H_u$ and
$H_d$ have expectation values, this gives, through equation
(\ref{componenty}),
masses for the quarks and leptons just as in the Standard Model.
If supersymmetry is unbroken, their scalar partners obtain
identical masses.  The last term is a supersymmetric
mass term for the Higgs fields.  Supersymmetry breaking,
essential to obtain a realistic model, will be discussed
momentarily.

The gauge symmetries actually permit many more couplings than
those written in eqn. (\ref{mssmw}).  Couplings such as $HL$, $\bar
u \bar d \bar d$, and others would violate baryon or lepton number
if they appeared.   Because these are dimension four, they are
unsuppressed (unless they have extremely tiny dimensionless
coefficients).  They can be forbidden by a symmetry, under which
ordinary fields are even (quarks, leptons, and Higgs bosons) while
there supersymmetric partners are odd.  This symmetry is called
R-parity.

By itself, this model is not realistic, since supersymmetry
is unbroken and all ordinary fields (quarks, leptons, gauge bosons, higgs)
are degenerate with their superpartners (squarks, sleptons,
gauginos).  The simplest solution to this is just to add
``soft breaking terms" which explicitly break the supersymmetry.
Because they are soft, they don't spoil the good features
of these theories.  These soft terms include mass terms
for the squarks and sleptons, Majorana mass terms
for the gauginos, and cubic couplings of the scalar fields,
\beq
\tilde m^2_{ij} \vert \phi^{i*} \phi^j \vert^2 + m_{\lambda} \lambda \lambda
+ m_A A_{ijk} \phi^i \phi^j \phi^k \label{softterms}
\eeq
In the minimal supersymmetric standard model, there are 105 such
couplings (counting real parameters).
We will think of all of these mass parameters as
being of order $m_{3/2}\sim m_Z$.
These parameters are highly constrained, both by low energy
physics (particularly by the suppression of flavor-changing
processes in weak interactions) and direct searches at LEP and the
Tevatron.
Theoretical approaches to understanding these soft breakings can
be divided broadly into two classes.  Both assume that some
dynamics gives rise to spontaneous breakdown of supersymmetry.
In ``gravity mediation," very high energy physics is responsible
for generating the soft terms; in gauge mediated models, lower
energy, gauge interactions communicate supersymmetry breaking to
ordinary fields.

R parity, if present, implies that the lightest of the new
particles, called the ``LSP" implied by supersymmetry is stable.  Typically
this is the partner of a neutral gauge or Higgs boson,  One can
calculate the abundance of these particles (neutralinos) as a function of
the various supersymmetry breaking parameters.
The assumption that the supersymmetry-breaking
masses are hundreds of GeV leads automatically
to a neutralino density of order
the dark matter density of the universe, and this particle
is a leading candidate for the dark matter.

\subsubsection{Baryogenesis in the MSSM and the NMSSM}
\label{sec:mssm}

Supersymmetric extensions of the Standard Model contain new
sources of CP
violation~\cite{dine_mssm1,dine_mssm2,huet_nelson} and an enlarged
set of parameters which allow a greater possibility of a
first-order transition~\cite{eqz,loops1,loops2,carena,cline-moore}.
So it would seem possible that electroweak baryogenesis could
operate effectively in these theories.

The new sources of CP violation may come, for example, from the chargino
mass matrix:
\begin{equation}
  \overline\psi_R M_\chi \psi_L = (\overline{\widetilde w^{^+}},\
  \overline{\widetilde h^{^+}_{2}} )_{R}
  \left(\begin{array}{cc}
             m_2 & g H_2(x) \\
           g H_1(x) & \mu
        \end{array}\right)
  \left(\begin{array}{c}
         \widetilde w^{^+} \\
         \widetilde h^{^+}_{1}
        \end{array}
  \right)_{\!\!L} + {\rm h.c.,}
\label{mp}
\end{equation}
where $\tilde{w}$ and $\tilde{h}$ are the superpartners of $W$-boson and
the charged Higgs,  and $m_2$ and $\mu$ are some mass parameters.

Other possible sources include phases in scalar masses (by field
redefinitions, some of these can be shifted from fermion to scalar
mass terms).  We will focus, however, on the terms in eq.~(\ref{mp}).

As long as $m_2$ and $\mu$ are complex, spatially varying phases in the
bubble wall provide a source of (spontaneous) CP
violation~\cite{lw1,lw2,ckn_spont}.  However, in light of the constraints
on Higgs and superpartner masses, the present window for electroweak
baryogenesis in the MSSM is very narrow if it exists at
all~\cite{cjk,carena_2002}.  As we discussed above, the lighter the Higgs
the easier it is to avoid the wash-out of baryon asymmetry produced in the
phase transition.  A light right-handed stop allows for a first-order phase
transition even for Higgs as heavy as $115$~GeV, which is barely consistent
with the current bounds.  The predictions of the light Higgs and the light
stop will soon be tested in experiment.

However, even for the most optimistic choice of parameters, it is difficult
to obtain a baryon asymmetry as large as the observed value quoted in
eq.~(\ref{baryondensity}).
Several parameters must be adjusted to maximize the baryon asymmetry.  In
particular, one must assume that the wall is very thin and choose the
``optimal'' bubble wall velocity $v_{\rm w} \approx 0.02$.
The origin of these difficulties lies, once again, in the strength of the
electroweak phase transition.  In the MSSM, the phase transition can be
enhanced if the right-handed stop (the scalar partner of the
top quark) is assumed to be very light, while the
left-handed stop is very heavy~\cite{carena}.
Then two-loop effects~\cite{loops1,loops2}
change the scalar potential sufficiently to allow for a first-order phase
transition; the lattice simulations support this perturbative
result~\cite{pt_mssm,pt_mssm1}.
However, severe constraints arise from the experimental bounds
on the chargino mass, as well as the chargino
contribution to the electric dipole moment of the
neutron~\cite{mssm_constraint,mssm_constraint1}.

\begin{figure}
\centerline{{\epsfysize=7cm \epsfbox{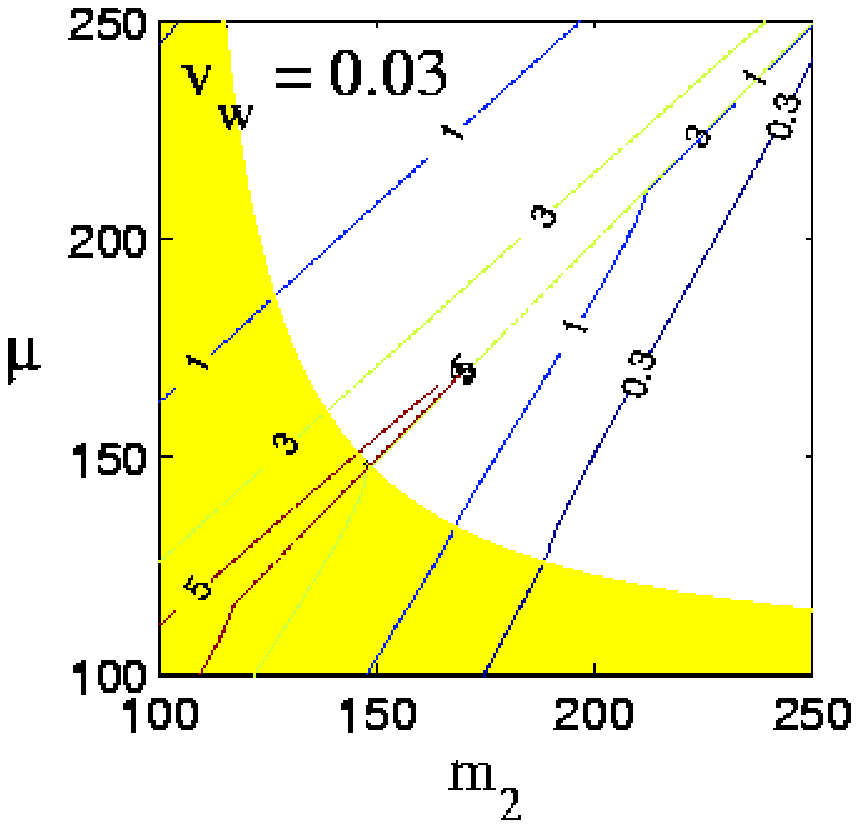}}
{  \epsfysize=7cm    \epsfbox{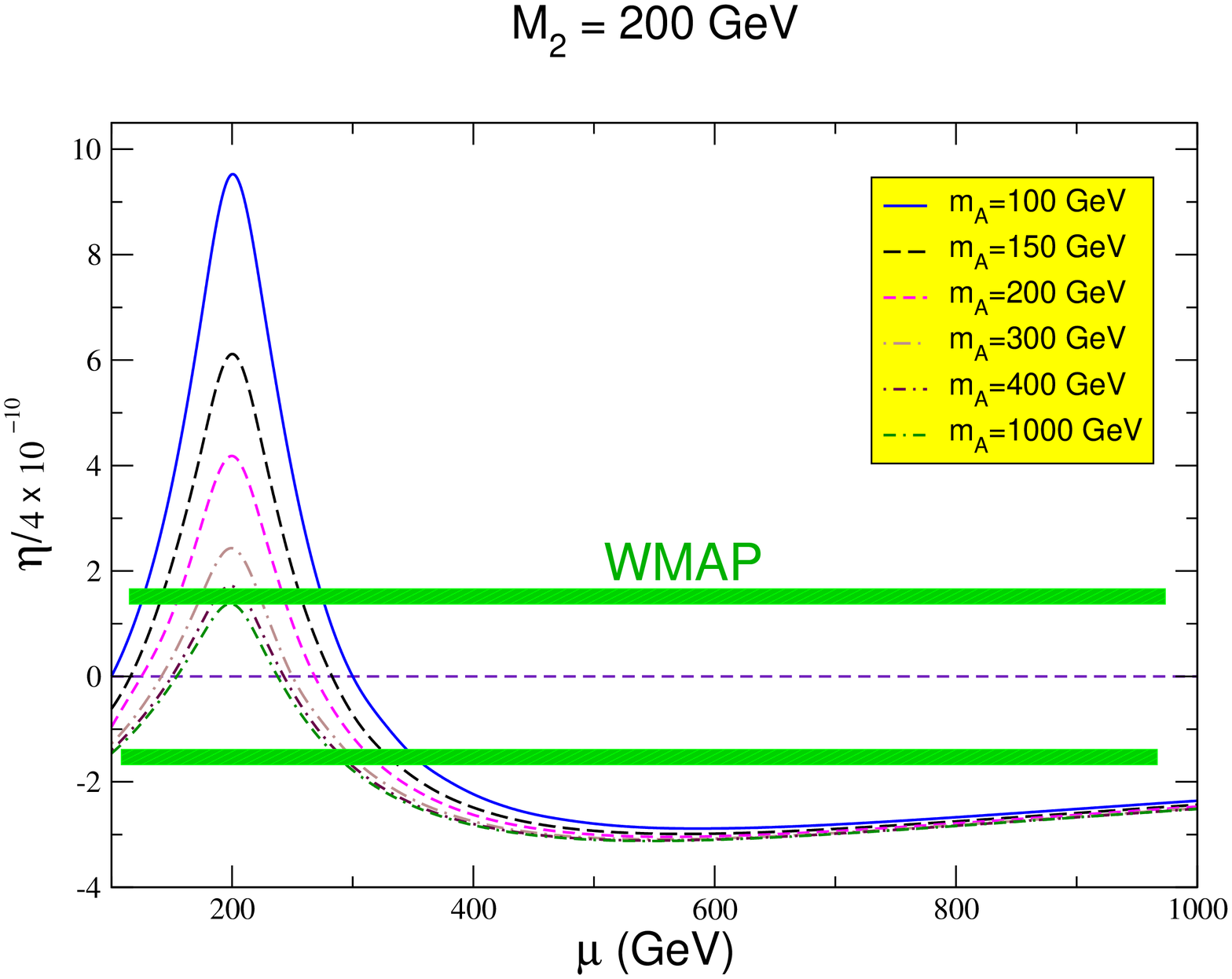} }
}
\caption{ Results of two independent calculations of baryon asymmetry
  in the MSSM.  On the left, contours of constant baryon asymmetry in units
  $10^{-10}$ calculated by \textcite{cjk} for the bubble wall velocity
  $v_{\rm w} = 0.03$ and $\tan \beta \lsim 3 $.  Mass units are
  GeV. Shaded regions are excluded by the LEP2 limit on the chargino
  mass, $m_{\chi^\pm}>104$ GeV. To maximize the baryon asymmetry, one
  assumes that the bubble wall is very narrow, $\ell_w \simeq 6/T$, and its
  velocity is $v_w=0.03$.  The
  plot on the right represents the results of \textcite{carena_2002} for
  {$\tan \beta =10$} and a maximal CP violating phase of the $\mu$
  parameter. The other parameters in both calculations have been chosen to
  maximize the resulting baryon asymmetry.  Also shown is the observed
  value of the baryon asymmetry reported by WMAP~\cite{map},
  {$\eta=6.1{\,}^{+0.3}_{-0.2}$}. }
\label{fig_mssm}
\end{figure}

Different calculations of the baryon asymmetry in the MSSM yield somewhat
different results~\cite{cjk,carena_2002}, as can be seen from
Fig.~\ref{fig_mssm}.  According to
\textcite{carena_2002}, it is possible to produce enough baryons if the
Higgs boson and the right-handed stop are both very light, near the present
experimental limits.   In any case, electroweak baryogenesis in the MSSM is
on the verge of being confirmed or ruled out by improving experimental
constraints~\cite{cline_review}.

The strength of the phase transition can be further enhanced by adding a
singlet Higgs to the model.  In the next-to-minimal supersymmetric model
(NMSSM), the phase transition can be more strongly
first-order~\cite{nmssm1,nmssm3,nmssm4}. The singlet also provides additional
sources of CP violation which
increase the baryon asymmetry~\cite{nmssm2}.

\subsubsection{Non-thermal electroweak baryogenesis at preheating}
\label{sec:preheat}

In light of these difficulties, various proposals have been put forth to
obtain a viable picture of electroweak baryogenesis.  These typically
involve more drastic departures from thermal equilibrium than the weakly
first order phase transitions described above.  The more extreme proposals
suppose that inflation occurred at the electroweak scale and kicked the
universe out of equilibrium, setting the stage for baryogenesis.  It is
generally believed that the natural scale for inflation is much higher than
$10^2$~GeV.  Although models with weak~\cite{weakscale_infl,grs} or
intermediate~\cite{supernatural} scale inflation have been
constructed, a
lower scale of inflation is generally difficult to reconcile with
the observed density perturbations $(\delta \rho/\rho)\sim 10^{-5}$.  As a
rule, the smaller the scale of inflation, the flatter the inflaton
potential must be to produce the same density fluctuations.   A weak-scale
inflation would require the inflaton potential to be extremely
flat, perhaps flatter than can plausibly be obtained in any
physical theory.

Of course, one does not have to assume that the same inflation is
responsible for $(\delta \rho/\rho)$ and for baryogenesis.  One could
imagine that the universe has undergone more than one inflationary period.
The primary inflation at a high scale could be responsible for the flatness
of the universe and for the observed density perturbations.  A
secondary inflation at the weak scale need not produce an
enormous expansion of the universe, and could create fertile soil for
baryogenesis.  One can 
debate the plausibility of invoking a second stage of inflation just for
this purpose.  In favor of such a possibility,
it has been argued that a low-scale
inflation might ameliorate the cosmological moduli problem
common to many supersymmetric theories~\cite{weakscale_infl,grs}.
Nevertheless, inflationary models at the electroweak scale, which would
help generate baryon asymmetry, usually suffer from naturalness
problems~\cite{lyth}, which may be less severe in some
cases~\cite{copeland}.  

What inhibits electroweak baryogenesis in the Standard Model is too much
equilibrium and too little CP violation.  Both of these problems
might be
rectified if inflation is followed by reheating to a temperature just below the
electroweak scale~\cite{kt,ggks}.  Reheating, especially its variant dubbed
{\em preheating}, involves a radical departure from thermal
equilibrium~\cite{kls}.

During inflation, all matter and radiation are inflated away.  When
inflation is over, the energy stored in the inflaton is converted to
thermal plasma.  There are several possibilities for this reheating
process.  One possibility is that the inflaton may decay perturbatively
into light particles, which eventually thermalize.  However, in a class
of models, a parametric resonance may greatly enhance the production of
particles in some specific energy bands~\cite{kls}.  This process, caused
by coherent oscillations of the inflaton, is known
as {\em preheating}.  Alternatively, the motion of the condensate may
become spatially inhomogenious on scales smaller than the horizon.  This
kind of transition in the motion of the inflaton, called spinodal
decomposition, may lead to a very rapid {\em tachyonic}
preheating~\cite{tachionic_preheating}.

All of these variants of reheating force the universe into a
non-equilibrium state after the end of inflation and before thermalization
takes place.  This is, obviously, an opportune time for baryogenesis.  The
usual considerations of sphaleron transitions do not apply to a
non-equilibrium system.  But it turns out that baryon-number violating
processes similar to sphaleron transitions do take place at
preheating~\cite{ggks,ck}, as well as tachyonic
preheating~\cite{smit,Garcia-Bellido:2003}.
This has been demonstrated by a combination of numerical and analytical
arguments.  In addition, preheating allows the coherent motions of some
condensates to serve as sources of CP violation~\cite{cgk}.  Such sources
are poorly constrained by experiment and could have significant impact on
baryogenesis.  It is conceivable, therefore, that the electroweak-scale
inflation could facilitate generation of the baryon asymmetry.

\subsection{Leptogenesis}
\label{sec:L}

Of the five scenarios for baryogenesis which we have listed in the
introduction, we have discussed two which are connected to very high energy
physics: Gravitational and GUT baryogenesis.  We have given cosmological
arguments why they are not likely.  These arguments depend on assumptions
which we cannot now reliably establish, so it is yet possible that these
mechanisms were operative.  But if we tentatively accept these arguments we
can significantly narrow our focus.  Similarly we have seen that
electroweak baryogenesis, while a beautiful idea, cannot be implemented in
the Standard Model, and probably not in its minimal supersymmetric
extension.  So again, while we can not rule out the possibility that
electroweak baryogenesis in some extension of the Standard Model is
relevant, it is tempting, for the moment, to view this possibility as
unlikely.  Adopting this point of view leaves leptogenesis and Affleck-Dine
baryogenesis as the two most promising possibilities.  What is exciting
about each of these is that, if they are operative, they have consequences
for experiments which will be performed at accelerators over the next few
years.

While there is no experimental evidence for supersymmetry apart from the
unification of couplings, in the last few years, evidence for neutrino
masses has become more and more compelling~\cite{nirreview}.  This comes
from several sources: the fact that the flux of solar neutrinos does not
match theoretical expectations, in the absence of masses and
mixings~\cite{superK_solar,sno}; the apparent observation of neutrino
oscillations among atmospheric neutrinos~\cite{superK_atmos}; and direct
measurements of neutrino mixing~\cite{kamland}.

We will not review all of these phenomena here, but just mention
that the atmospheric neutrino anomaly suggests oscillations
between the second and third generation of neutrinos:
\beq
\Delta m_{\nu}^2 = 10^{-2}-10^{-4} {\rm eV}^2
\eeq
with mixing of order one,
while the solar neutrino deficit suggests smaller masses
($\Delta m^2 \sim 10^{-6} {\rm eV}^2$).
There is other evidence for neutrino oscillation from accelerator
experiments.  The SNO experiment has recently provided persuasive
evidence in support of the hypothesis of mixing (as opposed to
modifications of the standard solar model).  The results from
Super-Kamiokande, SNO and KamLAND are in good agreement.  There is also
evidence of mixing from an experiment at Los Alamos (LSND).  This
result should be confirmed, or not, by the MiniBoone experiment at
Fermilab.  The mixing suggested by atmospheric neutrinos is
currently being searched for directly by accelerators.  The data
so far supports the mixing interpretation, but is not yet
decisive.

The most economical explanation of these facts is that neutrinos have
Majorana masses arising from lepton-number
violating dimension five operators.  (A Majorana mass is a mass for
a two component fermion, which is permitted if the fermion carries no
conserved charges.)  We have stressed that the leading operators
permitted by the symmetries
of the Standard Model which violate lepton number are non-renormalizable
operators of dimension
five, i.e. suppressed by one power of some large
mass.  Explicitly, these have the form:
\beq
{\cal L}_{lv}={1 \over M} L H L H
\eeq
Replacing the Higgs field by its expectation value $v$ gives a mass
for the neutrino of order $v^2 \over M$.
If $M= M_p$, this mass is too small to account for either set of
experimental results.  So one expects that some lower scale is
relevant.  The ``see-saw" mechanism provides a simple picture of
how this scale might arise.  One supposes that
in addition to the neutrinos of the Standard Model, there are some
$SU(2)\times U(1)$-singlet neutrinos, $N$.  Nothing forbids these
from obtaining a large mass.  This could be of order $M_{_{\rm GUT}}$,
for example, or a bit smaller.  These neutrinos could also couple
to the left handed doublets $\nu_{_L}$, just like right handed charged
leptons.  Assuming, for the moment, that these couplings are not
particularly small, one would obtain a mass matrix, in the $\{ N, \nu_{_L}
\}$ basis, of the form
\beq
M_{\nu} = \left ( \matrix{M & M_W \cr M_W^T & 0} \right )
\eeq
This matrix has an eigenvalue $M_W^2 \over M$.  The latter number is of the
order
needed to explain the neutrino anomaly for $M \sim 10^{13}$ or so,
i.e. not wildly different than the GUT scale and other scales
which have been proposed for new physics.

For leptogenesis~\cite{fy}, what is important in this model is that the
couplings of $N$ break lepton number.  $N$ is a heavy particle; it can
decay both to
$h+\nu$ and $h + \bar \nu$, for example.  The partial widths to each of
these final states need not be the same.  CP violation can enter through
phases in the Yukawa couplings and mass matrices of the $N$'s.  At tree
level, however, these phases will cancel out between decays to the various
states and their (would be) CP conjugates, as in the case
of GUTs we discussed earlier.  So it is necessary to consider
interference between tree and one loop diagrams with discontinuities, as in
Fig.~\ref{feynman}.  In a model with three $N$'s, there are CP-violating
phases in the Yukawa couplings of the $N$'s to the light Higgs,
  The
heaviest of the right handed neutrinos, say $N_1$, can decay to $\ell$ and
a Higgs, or to $\bar \ell$ and a Higgs.  At tree level, as in the
case of GUT baryogenesis, the rates for production of leptons and antileptons
are equal, even though there are CP violating phases in the couplings.
It is necessary, again, to look at quantum corrections, in which
dynamical phases can appear in the amplitudes.  At one loop, the decay
amplitude for $N$
has a discontinuity associated with the fact that the intermediate $N_1$
and $N_2$ can be on shell.  So one obtains an asymmetry
proportional to the imaginary parts of the Yukawa couplings of the
$N$'s to the Higgs:
\bea
\epsilon & = &
{\Gamma(N_1 \rightarrow \ell H_2)-\Gamma(N_1 \rightarrow \bar\ell \bar H_2)
\over \Gamma(N_1 \rightarrow \ell H_2) + \Gamma(N_1 \rightarrow \bar \ell
\bar H_2)} \\
& = & {1\over 8 \pi} {1 \over h h^\dagger}
\sum_{i=2,3} {\rm Im} [(h_{\nu} h_{\nu}^{\dagger})_{1i}]^2
f\left ({M_i^2 \over M_1^2} \right )
\eea
where $f$ is a function that represents radiative corrections.
For example, in the Standard Model $f=\sqrt{x} [ (x-2)/(x-1)+(x+1)
  \ln(1+1/x)]$, while in the MSSM $f=\sqrt{x} [ 2/(x-1)+\ln(1+1/x)]$.
Here we have allowed for the possibility of multiple
Higgs fields, with $H_2$ coupling to the leptons.
The rough order of magnitude here is readily understood by simply
counting loops factors.  It need not be terribly small.

Now, as the universe cools through temperatures of order the
of masses of the $N$'s, they drop out of equilibrium, and their
decays can lead to an excess of neutrinos over antineutrinos.
Detailed predictions can be obtained by integrating a suitable set
of Boltzman equations.   Alternatively, these particles can be produced out
of equilibrium, at prehearing following
inflation~\cite{Garcia-Bellido:2001cb}.

These decays, then, produce a net lepton number, but not baryon
number (and hence a net
$B-L$).
The resulting lepton number will be further processed by sphaleron
interactions, yielding a net lepton and baryon number (recall
that sphaleron interactions preserve $B-L$, but violate $B$ and
$L$ separately).  One can determine the resulting asymmetry by an
elementary thermodynamics exercise\cite{Harvey:1990qw}.
One introduces chemical
potentials for each neutrino, quark and charged lepton species.
One then considers the various interactions between the species at
equilibrium.  For any allowed chemical reaction, the sum of the
chemical potentials on each side of the reaction must be equal.
For neutrinos, the relations come from:
\begin{enumerate}
\item  the sphaleron interactions themselves:
\beq
\sum_i(3 \mu_{q_i} + \mu_{\ell_i}) =0
\eeq
\item  a similar relation for QCD sphalerons:
\beq
\sum_i(2 \mu_{q_i} - \mu_{u_i}-\mu_{d_i})=0.
\eeq
\item  vanishing of the total hypercharge of the universe:
\beq
\sum_i \left (\mu_{q_i} - 2 \mu_{\bar u_i} + \mu_{\bar d_i} - \mu_{\ell_i}
+\mu_{\bar e_i} \right ) + {2 \over N}\mu_H
=0
\eeq
\item  the quark and lepton Yukawa couplings give relations:
\beq
\mu_{q_i} - \mu_{\phi} - \mu_{d_j} = 0, \ \  \mu_{q_i} - \mu_{\phi} -
  \mu_{u_j} = 0,
\ \ \
\mu_{\ell_i} - \mu_{\phi} - \mu_{e_j}=0.
\eeq

\end{enumerate}

The number of equations here is the same as the number of
unknowns.
Combining these, one can solve for the chemical
potentials in terms of the lepton chemical potential, and finally
in terms of the initial $B-L$.  With N generations,
\beq
B= {8N+4 \over 22N + 13} (B-L).
\eeq

Reasonable values of the neutrino
parameters give asymmetries  of the order we seek
to explain.  Note sources of small numbers:
\begin{enumerate}
\item  The phases in the couplings
\item The loop
factor
\item  The small density of the $N$ particles
when they drop out of equilibrium.  Parametrically,
one has, e.g., for production,
\beq
\Gamma \sim e^{(-M/T)} g^2 T
\eeq
which is much less than $H \sim T^2/M_p$ once the density is
suppressed by $T/M_p$, i.e. of order $10^{-6}$ for a $10^{13}$
GeV particle.
\end{enumerate}

It is interesting to ask: assuming that these processes are the source of
the observed asymmetry, how many parameters which enter into the
computation can be measured, i.e.  can we relate the observed number to
microphysics.  It is likely that, over time, many of the parameters of
the light neutrino mass matrices, including possible CP-violating effects,
will be measured~\cite{nirreview}.  But while these measurements determine
some of the $N_i$ couplings and masses, they are not, in general, enough.
In order to give a precise
calculation, analogous to the calculations of nucleosynthesis, of the
baryon number density, one needs additional information about the masses of
the fields $N_i$~\cite{lepto_CP}.  One either requires some other
(currently unforseen) experimental access to this higher scale
physics, or a compelling theory of neutrino mass in which
symmetries, perhaps, reduce the number of parameters.

\subsection{Baryogenesis through Coherent Scalar Fields}

We have seen that supersymmetry introduces new
possibilities for electroweak baryogenesis.  But the most striking
feature of supersymmetric models, from the point of view of baryogenesis,
is the appearance of scalar fields carrying baryon and lepton number.
These scalars offer the possibility of coherent production of baryons.
In the limit that supersymmetry is unbroken, many of these scalars have
flat or nearly flat potentials.  They are thus easily displaced from
their minima in the
highly energetic environment of the early universe.  We will often refer to
such configurations as ``excited." Simple processes
can produce substantial amounts of baryon number.  This coherent production
of baryons, known as Affleck-Dine baryogenesis, is the focus of the rest
of this review.

\section{Affleck-Dine baryogenesis}
\label{sec:AD}

\subsection{Arguments for Coherent Production of the Baryon
Number}

In the previous section, we have reviewed several proposals for
generating the baryon number.  None can be firmly ruled out,
however all but two seem unlikely:  leptogenesis and Affleck-Dine
baryogenesis.  While the discovery of neutrino mass gives support
to the possibility of leptogenesis, there are a number of reasons
to consider coherent production:

\begin{itemize}
\item The Standard Model alone cannot explain the baryon asymmetry of the
  universe, the main obstacle being the heaviness of the Higgs.  One needs
  new physics for baryogenesis.  The requisite new physics may reside
  at a very high scale or at a lower scale.  An increasing body of evidence
  implies that inflation probably took place in the early universe.  Hence,
  baryogenesis must have happened at or after reheating.  To avoid
  overproducing weakly interacting light particles, for example gravitinos
  and other new states predicted in theory, one would like the reheat
  temperature not to exceed $10^9$~GeV.  This poses a problem for GUT
  baryogenesis.  This also limits possibilities for the leptogenesis.
  Affleck-Dine baryogenesis, on the other hand, is consistent with low
  energy and temperature scales required by inflation.

\item Supersymmetry is widely regarded as a plausible, elegant, and natural
  candidate for physics beyond the Standard Model.  Of the two simple
  scenarios for baryogenesis in the MSSM, the electroweak scenario is
  on the verge of being ruled out by accelerator constraints
  on supersymmetric particles, in sharp contrast with
  the AD scenario.

\item The remaining low-reheat SUSY scenario, Affleck-Dine baryogenesis,
  can naturally reproduce the observed baryon asymmetry of the universe.
  The formation of an AD condensate can occur quite generically in
  cosmological models.

\item The Affleck-Dine scenario potentially can give rise simultaneously to the
  ordinary matter and the dark matter in the universe.  This can explain
  why the amounts of luminous and dark matter are surprisingly close to
  each other, within one order of magnitude.  If the two entities formed in
  completely unrelated processes (for example, the baryon
  asymmetry
  from leptogenesis, while the dark matter from freeze-out of
  neutralinos), the observed relation $\Omega_{\rm DARK} \sim \Omega_{\rm
  matter}$ is fortuitous.\footnote{An additional {\em ad hoc} symmetry can
  also help relate the amounts of ordinary matter and dark
  matter~\cite{kaplan}.}

\item  Many particle physics models lead to significant production of
entropy at relatively late times~\cite{ckn_review}.
This dilutes whatever baryon
number existed previously.  Coherent production can be extremely
efficient, and in many models, it is precisely this
late dilution which yields the small baryon density observed
today.

In the rest of this section, we discuss Affleck-Dine baryogenesis
in some detail.

\end{itemize}

\subsection{Baryogenesis Through a Coherent Scalar Field:}

In supersymmetric theories, the ordinary quarks and
leptons are accompanied by scalar fields.  These scalar
fields carry baryon and lepton number.  A coherent field,
{\em i.e.}, a large classical value of such a field, can in
principle carry a large amount of baryon number.  As we will see, it
is quite plausible that such fields were excited in the
early universe.

To understand the basics of the mechanism, consider first
a model with a
single complex scalar field.  Take the lagrangian to be
\beq
{\cal L}= \vert \partial_{\mu} \phi \vert^2 -
m^2 \vert \phi \vert^2\label{lfree}
\eeq
This lagrangian has a symmetry, $\phi \rightarrow e^{i \alpha}
\phi$, and a corresponding conserved current, which we will
refer to as baryon number:
\beq
j^{\mu}_B=i(\phi^* \partial^{\mu} \phi -\phi \partial^{\mu}
\phi^*).\label{bcurrent}
\eeq
It also possesses a ``CP" symmetry:
\beq
\phi \leftrightarrow \phi^*.\label{simplecp}
\eeq
With supersymmetry in mind, we will think of $m$ as of order
$M_W$.

If we focus on the behavior of spatially constant fields,
$\phi(\vec x,t)=\phi(t)$, this system is equivalent to
an isotropic harmonic oscillator
in two dimensions.  This remains the
case if we add higher order terms which respect the
phase symmetry.  In supersymmetric models, however, we expect that
higher order terms will break the symmetry.  In the isotropic
oscillator
analogy, this corresponds to anharmonic terms which break the
rotational invariance.  With a general initial condition, the
system will develop some non-zero angular momentum.  If the motion
is damped, so that the amplitude of the oscillations decreases,
these rotationally non-invariant terms will become less important
with time.

Let us add interactions in the following way, which
will closely parallel what happens in the supersymmetric
case.  Include a set of quartic
couplings:
\beq
{\cal L}_I= \lambda \vert \phi \vert^4
+ \epsilon \phi^3 \phi^* + \delta \phi^4 + c.c.
\label{simplebviolation}
\eeq
These interactions clearly violate ``B".  For general complex
$\epsilon$ and $\delta$, they also violate $CP$.
In supersymmetric theories, as we will shortly see, the couplings
$\lambda,\epsilon,\delta\dots$ will be extremely small,
${\cal O}(M_W^2/M_p^2)$ or ${\cal O}(M_W^2/M_{_{\rm GUT}}^2)$.

In order that these tiny couplings lead to an appreciable baryon
number, it is necessary that the fields, at some stage, were very
large.
To see how the cosmic evolution of this system can lead
to a non-zero baryon number, first note that at very
early times, when the Hubble constant, $H \gg m$,
the mass of the field is irrelevant.
It is thus reasonable to suppose that at this
early time $\phi=\phi_o \gg 0$; later we will
describe some specific suggestions as to how
this might come about.  How does the field then evolve?
First ignore the quartic interactions.  In a gravitational
background, the equation of motion for the field is
\beq
D_{\mu}^2 \phi + {\partial V \over \partial \phi} = 0,
\label{covarianteqn}
\eeq
where $D_{\mu}$ is the covariant derivative.  For a spatially
homogeneous field, $\phi(t)$,
in a Robertson-Walker background, this becomes
\beq
\ddot \phi + 3 H \dot \phi + {\partial V \over \partial \phi}
=0.\label{rwscalar}
\eeq
At very early times, $H \gg m$, and so the system is highly
overdamped and essentially frozen
at $\phi_o$.  At this point, $B=0$.  However, once the universe
has aged enough that $H \ll m$, $\phi$ begins to oscillate.
Substituting
$H = { 1 \over 2 t}$ or
$H = {2 \over 3 t}$ for the radiation and matter dominated eras,
respectively, one finds
that
\beq
\phi = \left \{
\begin{array}{l}
{\phi_o \over (mt)^{3/2}} \sin(mt)~~{\rm
  (radiation)} \label{radsoln}
\\
{\phi_o \over (mt)}\sin(mt)~~{\rm (matter)}.\label{mattersoln}
\end{array} \right.
\eeq
In either case, the energy behaves, in terms of the scale
factor, $R(t)$, as
\beq
E \approx {m^2 \phi_o^2}({R_o\over R})^3\label{energydamping}
\eeq
i.e. it decreases like $R^3$, as would the energy of pressureless dust.
One can think of this oscillating field as a coherent state of $\phi$
particles with $\vec p=0$.

Now let's consider the effects of the quartic couplings.
Since the field amplitude damps with time,
their significance will decrease with time.
Suppose, initially,
that $\phi=\phi_o$ is real.  Then the imaginary part
of $\phi$ satisfies, in the approximation that $\epsilon$
and $\delta$ are small,
\beq
\ddot \phi_i + 3 H \dot \phi_i + m^2 \phi_i \approx {\rm Im}(\epsilon
+ \delta) \phi_r^3.\label{imaginarypart}
\eeq
For large times, the right hand falls as $t^{-9/2}$, whereas the left hand
side falls off only as $t^{-3/2}$.  As a result, just as in
our mechanical analogy, baryon number (angular momentum) violation
becomes negligible.  The
equation goes over to the free equation, with a solution
of the form
\beq
\phi_i = a_r{{\rm Im}(\epsilon + \delta) \phi_o^3
\over m^2 (mt)^{3/4}} {\rm sin}(mt + \delta_r)
~~{\rm (radiation),}
~~~~~\phi_i = a_m{{\rm Im}(\epsilon + \delta) \phi_o^3
\over m^3t} ~{\rm sin}(mt + \delta_m)~~{\rm (matter)},\label{imsolnmat}
\eeq
in the radiation and matter dominated
cases, respectively.  The constants
$\delta_m$, $\delta_4$, $a_m$ and $a_r$ can easily
be obtained numerically, and are of
order unity:
\beq
a_r = 0.85~~~~~a_m=0.85 ~~~~~ \delta_r= -0.91 ~~~~~
\delta_m = 1.54.\label{adeltavalues}
\eeq

But now we have a non-zero baryon number; substituting
in the expression for the current,
\beq
n_B= {2a_r {\rm Im}(\epsilon + \delta)} {\phi_o^2 \over m(mt)^2}
\sin(\delta_r + \pi/8) ~~{\rm (radiation)}
~~~~~n_B= {2a_m {\rm Im}(\epsilon + \delta)} {\phi_o^2 \over m(mt)^2}
\sin(\delta_m)~~{\rm (matter)}\label{nbmat}.
\eeq

Two features of these results
should be noted.  First, if $\epsilon$ and $\delta$
vanish, $n_B$ vanishes.  If they are real, and $\phi_o$
is real, $n_B$ vanishes.It is remarkable that the
lagrangian
parameters can be real, and yet $\phi_o$ can
be complex, still giving rise to a net baryon number.
We will discuss plausible initial values for the fields
later, after we have discussed supersymmetry breaking in the
early universe.  Finally, we
should point out that,
as expected, $n_B$ is conserved at late times.

This mechanism for generating baryon number could be
considered without supersymmetry.  In that case, it begs several questions:
\begin{itemize}
\item  What are the scalar fields carrying baryon number?
\item  Why are the $\phi^4$ terms so small?
\item  How are the scalars in the condensate
converted to more familiar particles?
\end{itemize}

In the context of supersymmetry, there is a natural answer to each of these
questions.  First, as we have stressed, there are scalar fields carrying
baryon and lepton number.  As we will see, in the limit that supersymmetry
is unbroken, there are typically directions in the field space in which the
quartic terms in the potential vanish.  Finally, the scalar quarks and
leptons will be able to decay (in a baryon and lepton number conserving
fashion) to ordinary quarks.

\subsection{Flat Directions and Baryogenesis}
\label{sec:flat}

To discuss the problem of baryon number generation,
we first want to examine the theory in a limit in which
we ignore the soft SUSY-breaking terms.  After all,
at very early times, $ H \gg M_W$, and these terms
are irrelevant.  We want to ask whether in a model
like the MSSM, some
fields can have large vev's, i.e. whether there are directions
in the field space for which the potential vanishes.   Before
considering the full MSSM, it is again helpful to
consider a simpler model, in this case a
theory with gauge group $U(1)$, and two chiral fields,
$\phi^+$ and $\phi^-$ with opposite charge.  We
take the superpotential simply to vanish.  In this
case the potential is
\beq
V={1 \over 2} D^2~~~~~D=g(\phi^{+*} \phi^+ -\phi^{-*}
\phi^-)\label{uoned}
\eeq
But $D$, and the potential, vanish if
$\phi^+=\phi^-=v$.
It is not difficult to work
out the spectrum in a vacuum of non-zero $v$.
One finds that there is one massless chiral field,
and a massive vector field containing a massive gauge
boson, a massive Dirac field, and a massive scalar.
\bigskip

Consider, now, a somewhat more elaborate example.
Let us take the MSSM and give expectation values to the Higgs and the
slepton fields of eqn.~(\ref{quantumnos}):
\beq
H_u= \left ( \matrix{0 \cr v} \right ) ~~~~~
L_1= \left ( \matrix{v \cr 0} \right ).\label{higgsflat}
\eeq
The $F$ term vanishes in this direction, since the potentially
problematic $H_u L$ term in the superpotential is absent by $R$
parity (the other possible contributions vanish because $Q=H_U=0$).
It is easy to see that the $D$-term for hypercharge
vanishes,
\beq
D_Y= g^{\prime~2}(\vert H_u \vert^2 - \vert L \vert ^2
 )=0.\label{vanishingdy}
\eeq
To see that the $D$ terms for $SU(2)$ vanishes,  one can work directly
with the Pauli matrices, or use, instead, the following
device which works for a general $SU(N)$ group.  Just
as one defines a matrix-valued gauge field,
\beq
(A_{\mu})^i_{~j}= A_{\mu}^a (T^a)^i_{~j},\label{matrixa}
\eeq
one defines
\beq
(D)^i_{~j}= D^a (T^a)^i_{~j}.\label{matrixd}
\eeq
Then, using the $SU(N)$ identity,
\beq
(T^a)^i_{~j}(T^a)^k_{~l}=\delta^i_{~l} \delta^k_{~j}-{1 \over N}
\delta^i_{~j} \delta^k_{~l}\label{sunidentity}
\eeq
the contribution to $(D)^i_{~j}$ from
a field, $\phi$, in the fundamental representation is simply
\beq
(D)^i_{~j}=\phi^{i*} \phi_j -{1 \over N} \vert \phi \vert^2
\delta^i_j. \label{dmatrix}
\eeq
In the present case, this becomes
\beq
(D)^i_{~j}=\left ( \matrix{\vert v \vert^2 & 0 \cr
0 & \vert v \vert^2} \right ) - {1 \over 2} \vert v \vert^2
\left ( \matrix{2 & 0 \cr
0 & 2} \right ) =0.\label{higgscancel}
\eeq

What is particularly interesting about this direction is that the field
carries a lepton number.  As we have seen, producing a lepton number is for
all intents and purposes like producing a baryon number.

Non-renormalizable, higher dimension terms, with more fields, can
lift the flat direction.  For example, the quartic term in the
superpotential:
\beq
{\cal L}_4 = {1 \over M}(H_u L)^2
\label{quarticl}
\eeq
respects all of the gauge symmetries and is invariant under
$R$-parity.  It gives rise to a potential
\beq
V_{\rm lift}= {\Phi^6 \over M^2}
\eeq
where $\Phi$ is the superfield whose vev parameterizes the flat
direction.

There are many more flat directions, and many
of these do carry baryon or lepton number.\footnote{The flat
  directions in the MSSM have been cataloged by~\textcite{flat}.}
A flat direction with both baryon and lepton number
excited is the following:
\beq
{\rm First ~generation:~}  Q^1_1=b ~~~~~\bar u_2=a
~~~~~L_2=b  \\
~~~~~{\rm Second}:~
\bar d_1
= \sqrt{\vert b \vert^2 + \vert a \vert^2} ~~~~~ {\rm Third}:~
\bar d_3=a.\label{thirdgen}
\eeq
(On $Q$, the upper index is a color index, the lower index
an $SU(2)$ index, and we have suppressed the generation
indices).

\smallskip
To check that this is indeed a flat direction, consider first
the $D$ terms for the various groups.  Using our earlier
matrix construction, we have:
\beq
SU(3):   \left ( \matrix{\vert b \vert^2 - (\vert a \vert^2
+\vert b \vert^2) & 0 & 0 \cr 0 & -\vert a \vert^2 & 0 \cr
0 & 0 & -\vert a \vert^2}\right ) - {\rm trace}=0
~~~~~SU(2):\left ( \matrix{\vert b \vert^2 & 0 \cr 0 &
\vert b \vert^2 } \right )- {\rm trace} =0
\eeq
\beq
U(1):  {1 \over 3} \vert b \vert^2 - {4 \over 3}
\vert a \vert^2 + {2 \over 3} \vert b \vert^2 + {2 \over 3} \vert a
\vert^2 + {2 \over 3} \vert a \vert^2 - \vert b \vert^2
= 0.\label{uoned1}
\eeq

The $F$ terms also vanish:
\beq
{\partial W \over \partial H_u} = \Gamma^{ab} Q^a \bar u^b = 0 ~~~~~{\rm
  and} \ \ {\partial W \over \partial H_d} = \gamma^{a}
Q^a \bar d^a=0
\label{hufterm}
\eeq
The first follows since the $\bar u$ fields have their
expectation values in different ``color slots"
than the $Q$ fields.
The second is automatically satisfied since the $\bar d$ and $Q$ fields have
expectation values in different generations, and these
Yukawa couplings don't mix generations.

Higher dimension operators again can lift this flat direction.
In this case the leading term is:
\beq
{\cal L}_{7}= {1 \over M^3}
[Q^1 \bar d^2L^1][\bar u^1 \bar d^2 \bar d^3].\label{lifts}
\eeq
Here the superscripts denote flavor.  We have suppressed color and
$SU(2)$ indices, but the braces indicate sets of fields which
are contracted in $SU(3)$ and $SU(2)$ invariant ways.
In addition to being completely gauge invariant, this operator
is invariant under ordinary $R$-parity.  (There are lower dimension
operators, including operators of dimension $4$, which
violate $R$-parity).
It gives rise to a term in the potential:
\beq
V_{\rm lift}={\Phi^{10} \over M^6}.\label{vflat}
\eeq
Here $\Phi$ refers in a generic way the fields whose vev's
parameterize the flat directions ($a$,$b$).

\subsection{Evolution of the Condensate}

For the cosmologies we wish to consider, despite the powers of $1\over M$,
these operators are quite important.  During inflation, for example, such
operators can determine the initial value of the field, $\Phi_o$ (here
$\Phi$ denotes in a generic way the fields which parameterize the flat
directions).

\subsubsection{Supersymmetry Breaking in the Early Universe}

We have indicated that higher dimension, supersymmetric operators
give rise to potentials in the flat directions.  To fully
understand the behavior of the fields in the early universe, we
need to consider supersymmetry breaking, which gives rise to
additional potential terms.

We have indicated, in equation~(\ref{softterms}), the sorts of
supersymmetry-breaking terms which we expect in supersymmetric theories.
In the early universe, we expect supersymmetry is much more badly broken.
For example, during inflation, the non-zero energy density (cosmological
constant) breaks supersymmetry.  Suppose that $I$ is the inflaton field,
and that the inflaton potential arises because of a non-zero value of the
auxiliary field for $I$, $F_I={\partial W \over \partial I}$ (see
eqn.~(\ref{auxiliary})).  $F_I$ is an order parameter for supersymmetry
breaking as are the auxiliary fields for any field; this quantity is
roughly constant during inflation. So, during inflation, supersymmetry is
broken by a large amount~\cite{drt_SUSYbreaking}.  Not surprisingly, as a
result, there can be an appreciable supersymmetry-breaking potential for
$\Phi$.  These contributions to the potential have the form:\footnote{When
supersymmetric theories are coupled to gravity, there are corrections to
eqn.~(\ref{potential}).  It also makes no sense to restrict the lagrangian
to be renormalizable.  Assuming that non-renormalizable couplings scale
with $M_p$, leads to eqn.~(\ref{correctionform}), as explained in
\cite{drt_SUSYbreaking}.}
\beq
V_H=H^2 \Phi^2 f(\Phi^2/M_p^2).\label{correctionform}
\eeq
It is perfectly possible for the second derivative
of the potential near the origin to be negative.  In this case, writing
our higher dimension term as:
\beq
W_n = {1 \over M^n} \Phi^{n+3}.\label{wschematic}
\eeq
the potential takes the form
\beq
V=-H^2 \vert \Phi \vert^2 + {1 \over M^{2n}}\vert
\Phi\vert^{2n+4}.
\eeq
The minimum of the potential then lies at:
\beq
\Phi_0 \approx M \left ({H \over M} \right
)^{1 \over n+1}.\label{phiohigherd}
\eeq
More generally, one can see that the higher the dimension of the
operator which raises the flat direction, the larger the starting
value of the field -- and the larger the ultimate value of the
baryon number.  Typically, there is plenty of time for the field
to find its minimum during inflation.  After inflation, $H$
decreases, and the field $\Phi$ evolves adiabatically,
oscillating slowly about the local minimum for some time.

Our examples illustrate that in models with R-parity,
the value of $n$, and hence the size of the initial field,
can vary appreciably.  With further symmetries,
it is possible that $n$ is larger,
and even that all operators which might lift the
flat direction are forbidden\cite{drt}.  For the rest of this
section we will continue to assume that the flat directions are
lifted by terms in the superpotential.  If they are not, the
required analysis is different, since the lifting of the flat
direction is entirely associated with supersymmetry breaking.

\subsubsection{Appearance of The Baryon Number}

The term in the potential, $\vert {\partial W \over \partial
\Phi}\vert^2$ does not break either baryon number or CP.  In most
models, it turns out that the leading sources of $B$ and $CP$
violation come from supersymmetry-breaking terms associated
with $F_I$.  These
have the form\footnote{Again, these arise from non-renormalizable
terms in the effective action, \cite{drt_SUSYbreaking}.}
\beq
a m_{3/2} W + b H W.\label{crucialterms}
\eeq
Here $a$ and $b$ are {\it complex}, dimensionless constants.  The
relative phase in these two terms, $\delta$, violates $CP$.  This
is crucial; if the two terms carry the same phase, then the
phase can be eliminated by a field redefinition, and we
have to look elsewhere for possible CP-violating effects.  Examining
equations (\ref{quarticl},\ref{lifts}), one sees that the term
proportional to $W$ violates $B$ and/or $L$.  In following the
evolution of the field $\Phi$, the important era occurs when
$H\sim m_{3/2}$.  At this point, the phase misalignment of the two
terms, along with the $B$ violating coupling, lead to the
appearance of a baryon number.  From the equations of motion,
the equation for the time rate of
change of the baryon number is
\beq
{dn_B \over dt}  =  {\sin(\delta)m_{3/2} \over M^n} \phi^{n+3}.
\eeq
Assuming that the relevant time is $H^{-1}$,
one is lead to the estimate (supported by numerical studies)
\beq
n_B= {1 \over M^n}\sin(\delta) \Phi_o^{n+3}.\label{roughb}
\eeq
Here, $\Phi_o$ is determined by $H \approx m_{3/2}$, i.e.
$\Phi_o^{2n+2}= m_{3/2}^2 M^{2n}$.

\subsection{The Fate of the Condensate}
\label{sec:fate}

Of course, we don't live in a universe dominated by a coherent scalar
field.  In this section, we consider the fate of a homogeneous
condensate, ignoring possible inhomogeneities.  The following
sections will deal with inhomogeneities, and the interesting array of
phenomena to which they might give rise.

We have seen that a coherent field can be thought of as a collection of zero
momentum particles.  These particles are long-lived,
since the particles to which they couple gain large
mass in the flat direction.  Were there no ambient plasma or other
fields, the condensate would eventually
decay.  However, there are a number of effects which lead the
condensate to disappear more rapidly, or to produce stable remnants.
Precisely what are the most important mechanisms depend on a number
of factors.  Perhaps most important is the rate of expansion, and the
dominant form of energy during this epoch.   The amplitude of
oscillations is also important.

It is impossible to survey all possibilities; indeed, it is likely
that all of the possibilities have not yet been imagined.
Instead, we
will adopt the picture
for inflation described in the previous section.  The features
of this picture are true of many models of inflation, but
by no means all.
We will suppose that the energy scale of inflation
is $E\sim 10^{15}$ GeV.  We assume that inflation
is due to a field, the inflaton $I$.  The amplitude of the
inflaton, just after inflation, we will take to be of order
 $M \approx
10^{18}$ GeV (the so-called reduced Planck mass).
Correspondingly, we will take the mass of the inflaton
to be $m_I = 10^{12}$ GeV
(so that $m_I^2 M_p^2 \approx E^4$).  Correspondingly, the
Hubble constant during inflation is of order $H_I \approx
E^2/M_p\approx 10^{12}$ GeV.

After inflation ends, the inflaton
oscillates about the minimum of its potential, much
like the field $\Phi$, until it decays.  We will suppose
that the inflaton couples to ordinary particles with a rate suppressed
by a single power of the Planck mass.  Dimensional analysis then
gives for the rough value of the
inflaton lifetime:
\beq
\Gamma_I  = {m_I^3 \over M^2} \sim 1 ~{\rm GeV}.\label{inflatonlifetime}
\eeq
The reheating temperature can then be obtained by equating the energy
density at time ($H \approx \Gamma$; $\rho = 3 H^2 M^2$  to the energy density
of the final plasma:~\cite{kolb}
\beq
T_R = T(t=\Gamma_I^{-1}) \sim (\Gamma_I M_p)^{1/2} \sim 10^9{\rm GeV}
\label{reheatinflaton}
\eeq

The decay of the inflaton is actually not sudden,
but leads to a gradual reheating of
the universe, as described, for example, in \cite{kolb}.  As a function of
time ($H$):
\beq
T \approx (T_R^2 H(t) M_p)^{1/4}.\label{inflatontemp}
\eeq

As for the field $\Phi$, our basic assumption is that during
inflation, it obtains a large value, in accord with equation
(\ref{phiohigherd}).  When inflation ends, the inflaton, by assumption
still dominates the energy density for a time, oscillating about
its minimum; the universe is matter dominated during this
period.
The field $\Phi$ now oscillates
about a time-dependent minimum, given by equation
(\ref{phiohigherd}).  The minimum decreases in value
with time, dropping to zero when
$H \sim m_{3/2}$.  During this evolution, a baryon number develops
classically.  This number is frozen once $H\sim m_{3/2}$.

Eventually the condensate will decay, through a variety of
processes.  As we have stressed, the condensate can be thought of
as a coherent state of $\Phi$ particles.  These particles --
linear combinations of the squark and slepton fields -- are
unstable and will decay.  However, for $H \le m_{3/2}$, the lifetimes
of these particles are much longer than in the absence of the
condensate.  The reason is
that the fields to which $\Phi$ couples have mass of order $\Phi$,
and $\Phi$ is large.  In most cases, the most important process
which destroys the condensate is what we might call evaporation:
particles in the ambient thermal bath can scatter off of the
particles in the condensate, leaving final states with only
ordinary particles.

We can make a crude estimate for the reaction rate as follows.
Because the particles which couple directly to $\Phi$ are heavy,
interactions of $\Phi$ with light particles must involve loops.
So we include a loop factor in the amplitude, of order
$\alpha_2^2$, the weak coupling squared.  Because of the large
masses, the amplitude is suppressed by $\Phi$.  Finally, we need
to square and multiply by the thermal density of scattered
particles.  This gives:
\beq
\Gamma_p \sim \alpha_2^2 \pi {1 \over \Phi^2}(T_R^2 H
M)^{3/4}.\label{condensateraction}
\eeq
The condensate will evaporate when this quantity is
of order $H$.  Since we know the time dependence of $\Phi$, this allows us
to solve for this time.  One finds that equality occurs, in the case $n=1$,
for $H_I \sim 10^2-10^3$~GeV.  For $n>1$, it occurs significantly later
(for $n< 4$, it occurs before the decay of the inflaton; for $n \ge 4$, a
slightly different analysis is required than that which follows).  In other
words, for the case $n=1$, the condensate evaporates shortly after the
baryon number is created (but for more complications, see below), but for
larger $n$, it evaporates significantly later.

The expansion of the universe is unaffected by the condensate as long as
the energy density in the condensate, $\rho_\Phi \sim m_\Phi^2 \Phi^2$, is
much smaller than that of the inflaton, $\rho_I \sim H^2 M^2$.  Assuming
that $m_\Phi \sim m_{3/2}\sim 0.1-1$~TeV, a typical supersymmetry breaking
scale, one can estimate the ratio of the two densities at the time when
$H\sim m_{3/2}$ as
\beq
\frac{\rho_\Phi}{\rho_I} \sim \left ( \frac{m_{3/2}}{M_p} \right
)^{2/(n+1)}.
\label{baryontoinflaton}
\eeq

We are now in a position to calculate the
baryon to photon ratio in this model.  Given our estimate of the
inflaton lifetime, the coherent motion of the inflaton still
dominates the energy density when the condensate evaporates.
The baryon number is just the $\Phi$ density just before
evaporation divided by the $\Phi$ mass (assumed of order $m_{3/2}$),
while the inflaton number is $\rho_I/M_I$.  So the
baryon to inflaton ratio follows from eqn.~(\ref{baryontoinflaton}).
With the assumption that the inflaton
energy density is converted to radiation at the reheating
temperature, $T_R$, we obtain:
\beq
\frac{n_B}{ n_{\gamma}} \sim
\frac{n_B}{(\rho_I/T_R)} \sim \frac{n_B}{n_\Phi} \frac{T_R}{m_\Phi}
\frac{\rho_\Phi}{\rho_I} \sim
10^{-10} \left (\frac{T_R}{10^9 {\rm GeV} }
\right ) \, \left ( \frac{M_p}{m_{3/2}}
\right )^{\frac{(n-1)}{(n+1)}}.
\eeq

Clearly the precise result depends on factors beyond those
indicated here explicitly, such as the precise mass of the $\Phi$
particle(s).  But as a rough estimate, it is rather robust.
For $n=1$, it is in {\it precisely the right range} to explain the observed
baryon asymmetry.  For larger $n$, it can be significantly larger.  While
this may seem disturbing, it is potentially a significant
virtue.  Many
supersymmetric models lead to creation of entropy at late times.  For
example, in string theory one expects the existence of other ``light"
($m\approx m_{3/2}$) fields, known as ``moduli."  These fields lead to
cosmological difficulties~\cite{moduli}, unless, when they decay, they
reheat the universe to temperatures of order $10$ MeV, after which
nucleosynthesis can occur.  These decays produce a huge amount of entropy,
typically increasing the energy of the universe by a factor of $10^7$.  The
baryon density is diluted by a corresponding factor.  So in such a
picture, it is necessary that the baryon number, prior to the moduli decay,
should be of order $10^{-3}$.  This is not the only cosmological model
which requires such a large baryon number density.

There are many issues in the evolution of the condensate which we
have not touched upon.  One of the most serious is related to
interactions with the thermal bath~\cite{anisimov_dine,campbell_ellis}.
In the case
$n=1$, $\Phi_o$ is not so large, and, while the particles which
$\Phi$ couples to get mass of order $\Phi_o$, they may be in thermal
equilibrium.  In this case, the $\Phi$ particles decay much
earlier.  This typically leads to significant suppression of the
asymmetry, and the viability of the AD mechanism depends on the
precise values of the parameters.

Overall, then, there is a broad range of parameters for which the
AD mechanism can generate a value for $n_B \over n_{\gamma}$ equal
to or larger than that observed.  This baryon number is generated
long after inflation, so inflationary reheating does not provide
any significant constraint.  It can be large, allowing for
processes which might generate entropy rather late.

\subsection{Inhomogeneities and the Condensate}

We have so far assumed that the condensate is homogeneous.  But,
as we will now show, under some circumstances the condensate is
unstable to fragmentation.  This appears to be related to another
feature of theories with scalars:  the possible existence of
non-topological solitons.  These can alter the picture of baryon
number generation, and could conceivably be a dark matter
candidate.  This is the subject of this section.

\subsubsection{Stability and fragmentation}

To analyze the stability of the condensate~\cite{ks}, we write the complex
field $\phi=\rho e^{i\Omega}$ in terms of its radial component and a phase,
both real functions of space-time.  We are interested in the evolution of
the scalar field in the small-VEV domain, where the baryon number violating
processes are suppressed, and we will assume that the scalar potential
preserves the U(1) symmetry: $U(\vf)=U(\rho)$, where $U(\rho)$ may depend
on time explicitly. The classical equations of motion in the spherically
symmetric metric $ds^2=dt^2-a(t)^2 dr^2$ are

\begin{eqnarray}
\ddot \Omega +3 H \dot \Omega - \frac{1}{a^2(t)} \Delta \Omega +\frac{2
\dot
  \rho}{\rho}
\dot \Omega - \frac{2}{a^2(t) \rho} (\dd_i \Omega)(\dd^i \rho) & = & 0,
\label{eqnmtn1}
\\
\nonumber \\
\ddot \rho + 3 H\dot \rho - \frac{1}{a^2(t)} \Delta \rho -
\dot \Omega^2 \rho + \frac{1}{a^2(t)} (\dd_i \Omega)^2
\rho + (\dd U/\dd \rho) & = & 0,
\label{eqnmtn2}
\end{eqnarray}
where dots denote time derivatives, and the space coordinates are
labeled by the Latin indices that run from 1 to 3. The Hubble
constant, again, is
$H=\dot a/a$, where $a(t)$ is the scale factor; it is equal to $t^{-2/3}$
or $t^{-1/2}$ for a matter or radiation dominated universe,
respectively.

{}From the equations of motion (\ref{eqnmtn1}) and (\ref{eqnmtn2}), one
can derive the equations for small perturbations $\delta \Omega$ and $
\delta \rho$:

\begin{eqnarray}
\ddot{\delta \Omega} + 3 H \dot{(\delta \Omega)}
- \frac{1}{a^2(t)} \Delta (\delta \Omega) +\frac{2 \dot
  \rho}{\rho} \dot{(\delta \Omega)}+ \frac{2 \dot \Omega }{\rho}
\dot{(\delta \rho)}
- \frac{2\dot \rho \dot \Omega }{\rho^2} \delta \rho
 & = & 0, \label{eqndelta1} \\
\nonumber \\
\ddot{\delta \rho} + 3 H \dot{(\delta \rho)}
- \frac{1}{a^2(t)} \Delta (\delta \rho)
-2 \rho \dot \Omega \dot{(\delta \Omega)}+ U'' \delta \rho - \dot \Omega^2
\delta \rho & = & 0.
\label{eqndelta2}
\end{eqnarray}

To examine the stability of a homogeneous solution $\vf(x,t)=\vf(t)
\equiv \rho (t) e^{i \Omega(t)}$, let us consider a perturbation $\delta
\rho , \delta \Omega \propto e^{S(t) - i \vec{k}\vec{x}} $ and look for
growing modes, ${\rm Re} \, \alpha>0$, where $\alpha = dS/dt $. The
value of $k$ is the spectral index in the comoving frame and is
red-shifted with respect to the physical wavenumber $\tilde{k}=k/a(t)$
in the expanding background. Of course, if an instability develops,
the linear approximation is no longer valid. However, we assume that
the wavelength of the fastest-growing mode sets the scale for the high
and low density domains that eventually evolve into Q-balls. This
assumption can be verified {\it post factum} by comparison with a
numerical solution of the corresponding partial differential equations
(\ref{eqnmtn1}) and (\ref{eqnmtn2}), where both large and small
perturbations are taken into account.

The dispersion relation follows from the equations of motion:

\beq
\left [
\alpha^2+ 3H\alpha +\frac{k^2}{a^2} + \frac{2 \dot \rho}{\rho} \alpha \right
]
\left [ \alpha^2+ 3H\alpha +\frac{k^2}{a^2} - \dot \Omega^2+U''(\rho)
\right ]+
4 \dot \Omega^2 \left [ \alpha - \frac{\dot \rho}{\rho} \right ] \alpha = 0.
\label{dr}
\eeq

If $(\dot \Omega^2-U''( \rho ))>0$, there is a band of growing modes that
lies between the two zeros of $\alpha(k)$, $0<k<k_{\max}$, where

\beq
k_{\max}(t)=
a(t) \sqrt{\dot \Omega^2-U''( \rho )}.
\label{band}
\eeq

This simple linear analysis shows that when the
condensate is ``overloaded'' with charge, that is when
$\omega(t)=\dot{\Omega}$ is larger than the second derivative of the
potential, an instability sets in.
Depending on how  $k_{\max}(t)$, defined by relation (\ref{band}), varies
with time, the modes in the bands of instability may or may not have time
to develop fully.

\begin{figure}
\setlength{\epsfxsize}{3.3in}
\centerline{\epsfbox{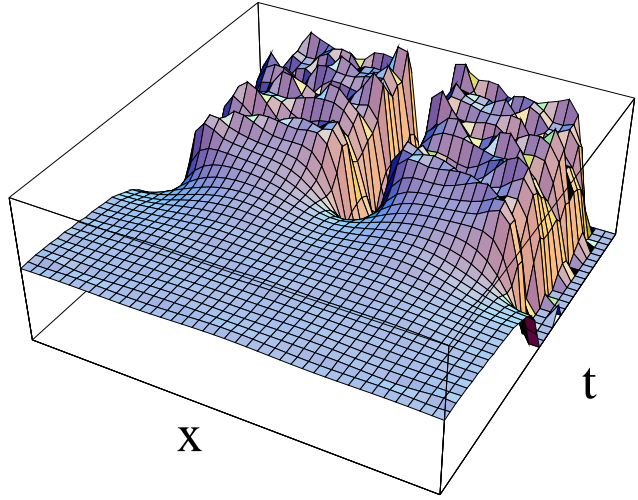}}
\caption{
The charge density per comoving volume in (1+1) dimensions for a sample
potential analyzed numerically during the period when the spatially
homogeneous condensate breaks up into high- and low-density domains.
Two domains with high charge density are expected to form Q-balls.
}
\label{fig_charge}
\end{figure}

Numerical analyses~\cite{kasuya1,kasuya2,kasuya3,enqvist_num}, which can
trace the evolution of unstable modes beyond the linear regime, have shown
that fragmentation of the condensate is a generic phenomenon.  Numerically
one can also study the stability of rapidly changing solutions, hence
relaxing the adiabaticity condition assumed above.  This aspect is relevant
to the cases where the baryon number density is small and the radial
component of the condensate, $\rho(t)$, exhibits an oscillatory behavior
changing significantly on small time scales.  An interesting feature of
this non-adiabatic regime is that both baryon and anti-baryon lumps may
form as a result of fragmentation~\cite{enqvist_num}.

\subsubsection{Lumps of scalar condensate: Q-Balls}
\label{sec:q}

Perhaps the most familiar soliton solutions of non-linear field
theories, such as magnetic monopoles and vortices, can be
uncovered by topological arguments.  However,
field theories with scalar fields often admit non-topological
solitons~\cite{coleman,fls,rosen}, Q-balls,
which may be stable or may decay into fermions~\cite{ccgm}.
Q-balls appear when a
complex scalar field $\vf$ carries a conserved charge with respect to some
global U(1) symmetry.  In supersymmetric generalizations of the Standard
Model, squarks and sleptons, which carry the conserved baryon and
lepton numbers, can form Q-balls.


Let us consider a field theory with a scalar potential $U(\vf) $
which has a global minimum $U(0)=0$ at $\vf=0$.
Let $U(\vf)$ have an unbroken global U(1) symmetry
at the global minimum: $\vf \rightarrow \exp\{i \theta \} \vf $.
We will look for solutions of the classical equations by
minimizing the energy
\beq
E=\int d^3x \ \left [ \frac{1}{2} |\dot{\vf}|^2+
\frac{1}{2} |\nabla \vf|^2
+U(\vf) \right]
\label{e}
\eeq
subject to the constraint that the configuration has a definite
charge, $Q$,
\beq
Q= \frac{1}{2i} \int \vf^{*} \stackrel{\leftrightarrow}{\partial}_t
\vf \, d^3x
\label{Qt}
\eeq

To describe the essential features of Q-balls in a simple way, we will,
following~\textcite{coleman}, use a thin wall ansatz for the Q-ball

\beq
\phi(x,t) = e^{i \omega t} \bar{\phi}(x),
\label{ansatz1}
\eeq
where

\beq
\bar{\phi}(x) = \left \{
\begin{array}{ll}
0, & \sqrt{\vec{x}^2} > R \\
\phi_0 & \sqrt{\vec{x}^2} \leq R
\end{array}
\right.
\label{ansatz2}
\eeq
(for the real solution, the field varies rapidly between the two
regions, changing on a scale of order the Compton wavelength of
the $\phi$ particle).

Assuming that $Q$ is large, let us neglect the gradient terms (relevant
only for the wall energy).  Then the global charge and the energy
of the field configuration (\ref{ansatz1},\ref{ansatz2}) are given by

\beq
Q=\omega \phi_0^2 V,
\eeq
where $V=(4/3)\pi R^3$, and

\beq
E= \frac{1}{2} \omega^2 \phi_0^2 V + U(\phi_0) V = \frac{1}{2} \frac{Q^2}{V
  \phi_0^2} + V U(\phi_0)
\eeq

We now minimize $E$ with respect to $V$, obtaining

\beq
E=\sqrt{\frac{2 U(\phi_0)}{\phi_0^2}}
\eeq
It remains to minimize the energy with respect to variations of $\phi_0$.
A non-trivial minimum exists as long as

\beq
U(\vf) \left/ \vf^2 \right. = {\rm min},
\ \ {\rm for} \
\vf=\vf_0>0;
\label{condmin1}
\eeq
if this condition is satisfied, a Q-ball solution exists.

So far we have
assumed a particular ansatz, neglected the gradient terms, {\em etc.}
These assumptions can be avoided in a slightly
more involved derivation~\cite{ak_qb} using the method of Lagrange
multipliers.  We want to minimize

\beq
\e = E+  \omega \left [ Q- \frac{1}{2i} \int \vf^* \dd_t \vf \, d^3x
\right ],
\label{Ew1}
\eeq
where $ \omega$ is a Lagrange multiplier.  (It is no accident that we
use the same letter, $\omega$.  The value of the Lagrange multiplier at the
minimum will turn out to be equal to the time derivative of the phase.)
Variations of $\vf(x,t)$ and those of $\omega$ can now be treated
independently, the usual advantage of the Lagrange method.

One can re-write equation (\ref{Ew1}) as

\beq
\e = \int d^3x \, \frac{1}{2} \left | \frac{\dd}{dt} \vf
- i \omega  \vf
\right |^2 \ + \ \int d^3x \, \left [\frac{1}{2}  |\nabla \vf |^2
+ \hat{U}_\omega(\vf)
\right ] + \omega Q ,
\label{Ewt}
\eeq
where

\beq
\hat{U}_\omega (\vf) = U(\vf)\ - \ \frac{1}{2} \, \omega^2 \, \vf^2.
\label{Uhat}
\eeq

We are looking for a solution that extremizes $\e$, while all the physical
quantities, including the energy, $E$,  are time-independent.  Only the
first term in equation (\ref{Ewt}) appears to depend on time explicitly,
but it is positive definite and, hence, it should vanish at the minimum.
To minimize this contribution to the energy, one must choose, therefore,

\beq
\vf (x,t) = e^{i\omega t} \vf (x),
\label{tsol}
\eeq
where $\vf(x)$ is real and independent of time.  We have thus
derived
equation (\ref{ansatz1}).  For this solution, equation (\ref{Qt}) yields

\beq
Q= \omega \int \vf^2(x) \ d^3x
\label{Qw}
\eeq

It remains to find an extremum of the functional

\beq
\e = \int d^3x \, \left [\frac{1}{2} |\nabla \vf(x) |^2
+ \hat{U}_\omega(\vf(x))
\right ] + \omega Q ,
\label{Ew}
\eeq
with respect to $\omega$ and the variations of $\vf(x)$ independently.
We can first minimize $\e$ for a fixed $\omega$, while varying the shape of
$\vf(x)$.  If this were an actual potential for a scalar field in three
dimensions, one would have the possibility of tunneling between the zero
energy configuration at the origin and possible lower energy configurations
at non-zero $\phi$ (Fig. 6).  Tunneling, in the semiclassical
approximation, is described by the bounce, $\bar{\vf}_\omega (x)$, the
solution of the classical equations which asymptotes to the ``false vacuum"
at the origin~\cite{tunn1}.
The first
term in equation (\ref{Ew}) is then nothing but the three-dimensional
Euclidean action $S_3 [\bar{\vf}_\omega (x) ]$ of this bounce solution.
This is a very useful correspondence.  In particular, the condition
for the existence of
solution is simply a corollary: as long as $\hat{U}_\omega (\vf)$ has a
minimum below zero, the bounce exists, and so does the Q-ball, $\vf
(x,t)=\exp \{i \omega t \} \bar{\vf}(x)$.

\begin{figure}
\setlength{\epsfxsize}{3.3in}
\centerline{\epsfbox{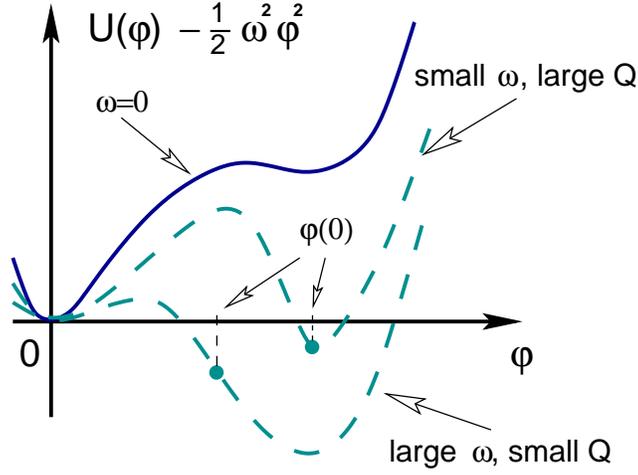}}
\caption{Finding a Q-ball is equivalent to finding a bounce that describes
tunneling in the potential {$\hat{U}_\omega(\vf)= U(\vf)-(1/2) \omega^2
\vf^2 $}.
The thin-wall approximation is good for large {$Q$} (upper dashed
line), but breaks down when {$Q$} is small and, therefore, $\omega$ is
almost as large as the mass term at the origin.  In the latter case (lower
dashed line), the ``escape point'', {$\vf(0)$} is close to the zero of the
potential and is far from the global minimum.
}
\label{fig1}
\end{figure}

The bounce, and hence the Q-ball, exist if there exists a value of $\omega
$, for which the potential $\hat{U}_\omega(\vf)$ has both a local minimum
at $\vf =0$ and a global minimum at some other value of $\vf$.  This
condition can be re-phrased without reference to $\omega$: a Q-ball
solution exists
if

\beq
U(\vf) \left/ \vf^2 \right. = {\rm min},
\ \ {\rm for} \
\vf=\vf_0>0
\label{condmin}
\eeq
The corresponding effective potential
$\hat{U}_{\omega_0} (\vf)$, where $\omega_0=\sqrt{2 U(\vf_0^2)/\vf^2}$,
has two degenerate minima, at $\vf=0$ and $\vf=\vf_0$.
The existence of the bounce solution $\bar{\vf}_\omega (x)$ for
$\omega_0<\omega < U''(0)$ follows \cite{tunn1,cgm} from the fact that
$\hat{U}_\omega (\vf)$ has a negative global minimum in addition to the
local minimum at the origin.  \textcite{cgm} also showed that the solution
is spherically symmetric: $\bar{\vf}(x)=\bar{\vf}(r), \
r=\sqrt{\vec{x}^2}$.

The soliton we want to construct is precisely this bounce for the right
choice of $\omega$, namely that which minimizes $\e$.
The last step is to find an extremum of

\beq
\e= S_3 [\bar{\vf}_\omega (x)] + \omega Q
\label{Ebounce}
\eeq
with respect to $\omega$.  One can prove the existence of such an
extremum~\cite{ak_qb}. Finally, the soliton is  of the
form (\ref{tsol}), with  $\omega$ that minimizes $\e$ in eq.~(\ref{Ebounce}).

Having obtained the solution, one can compute its energy (mass). For a
finite $\vf_0$ in eq.~(~\ref{condmin}), in the limit of large $Q$, the
Q-ball has a thin wall, and its mass is given by

\beq
M(Q)= \omega_0 Q
\label{MQthin}
\eeq

Supersymmetric generalizations of the Standard Model have scalar potentials
with flat directions lifted only by supersymmetry breaking terms.  Q-balls
may form with a light scalar field $\vf$ that corresponds to that flat
direction.  If the potential $V(\vf)=\mu^4={\rm const }$ for large $\vf$,
then the minimum in eq.~(\ref{condmin}) is achieved for $\vf_0 = \infty $.
In this case,
\beq
M (Q) = \mu Q^{3/4}.
\label{MQflat}
\eeq

More generally, if the potential grows slower than $\vf^2$, {\em i.e.}
$V(\vf)\propto \vf^p, p<2$, condition~(\ref{condmin}) is not satisfied at
any finite value of $\phi_0$, and

\beq
M(Q) \sim \mu \, Q^{(3-p/2)/(4-p)}.
\eeq

It is important in what follows that the mass per unit charge is not
a constant, but is a decreasing function of the total global charge $Q$.
There is a simple reason why the soliton mass is not proportional to
$Q$.  Since $U(\vf)/\vf^2$ has no minimum, the scalar VEV can extend as far
as the derivative terms allow it.  When the next unit of charge is added,
the Q-ball increases in size, which allows the scalar VEV to increase as
well.  Hence, the larger the charge, the greater is the VEV, and the
smaller is energy per unit charge.

\subsubsection{ AD Q-balls}

The condition (\ref{condmin}) suggests that slowly growing
potentials, of the sort that arise in the flat directions of the
MSSM, are a likely place to find Q-balls.
Q-balls can develop along flat directions that carry non-zero baryon
number, lepton number, or both.  Each flat direction can be parameterized
by a gauge-invariant field, carrying these global
quantum numbers.   So the discussion of gauge singlet
fields of the previous section also applies to baryonic and leptonic Q-balls
in the MSSM.  This statement may seem surprising, since
all scalar baryons in the MSSM transform
non-trivially under the gauge group.
Although scalars with gauge
interactions can also make Q-balls~\cite{gauged_Q}, in the case of the MSSM
the color structure of large Q-balls is rather simple~\cite{kst}.  If a
Q-ball VEV points along a flat direction, its scalar constituents form a
colorless combination (otherwise, that direction would not be flat because
of non-vanishing D-terms).

In one proposal for the origin
of supersymmetry breaking,
``gauge-mediated" supersymmetry breaking,
breaking, the flat directions are
lifted by potentials which grow quadratically for small values of the
fields, and then level off to a logarithmic plateau at larger $\phi$.
Q-balls in such a potential have masses given by eq.~(\ref{MQflat}).  In
another proposal,
``gravity-mediated" scenarios, the potentials which arise from supersymmetry
breaking grow roughly quadratically even for very large VEV.  Whether
Q-balls exist is thus a detailed, model-dependent question.  Q-balls in
these potentials have masses proportional to the first power of $Q$.

By construction, Q-balls are stable with respect to decay into scalars.
However, they can decay by emitting fermions~\cite{ccgm}. If the Q-ball has
zero baryon number, it can decay by emitting light neutrinos~\cite{ccgm}.

However, if a baryonic Q-ball (``B-ball'') develops along a flat direction,
it can also be stable with respect to decay into fermions.  Stability
requires that the baryon number be large enough. A Q-ball
with baryon number $Q_B$ and mass $M(Q_B)$ is stable if its mass is below
the mass of $Q_B$ separated baryons.  For a Q-ball in a flat potential of
height $M_S$, the mass per unit baryon number
\beq
\frac{M(Q_B)}{Q_B} \sim M_S Q^{-1/4}
\eeq
Models of gauge-mediated supersymmetry breaking produce flat potentials
with $M_S\sim 1-10~{\rm TeV}$.
If the mass per baryon number is less than the proton mass, $m_p$,
then the
Q-ball is entirely stable because it does not have enough energy to decay
into a collection of nucleons with the same baryon number.  This condition
translates into a lower bound on $Q_B$:
\beq
\frac{M(Q_B)}{Q_B}
\ <  1 {\rm GeV} \
\Rightarrow
Q_B \gg \left (\frac{M_S}{ 1 {\rm GeV}}
\right )^4 \gsim 10^{16}.
\eeq

\subsubsection{Dark matter in the form of stable B-balls}
\label{dark_stable}

Stable Q-balls that form from an AD condensate are a viable candidate for
dark matter.  Even if they are
unstable, their decay can produce neutralinos at late times, when these
neutralinos are out of equilibrium.  One way or another, some dark matter
can arise from AD baryogenesis.

Moreover, since both the ordinary matter and the dark matter have the same
origin in the AD scenario, one can try to explain why their amounts in the
universe are fairly close~\cite{ls,jedamzik,em_d,fy_d}.

Since the MSSM with gauge-mediated supersymmetry breaking contains stable
objects, baryonic Q-balls, it is natural to ask whether they can constitute
the dark matter.  Stable Q-balls can be copiously produced in the course of
AD baryogenesis.  Of the dark-matter candidates that have been considered,
most were weakly interacting particles, and for good reason.  If
dark-matter particles have strong ({\em relative to their mass})
interactions with matter, these interactions might facilitate their loss of
momentum and angular momentum, forcing
them into the galactic disks, along with ordinary matter.  But
astronomical observations show that
the dark matter forms spherical halos about galaxies,
not disks.

\begin{figure}
\setlength{\epsfxsize}{4.5in}
\epsfclipon
\centerline{\epsfbox{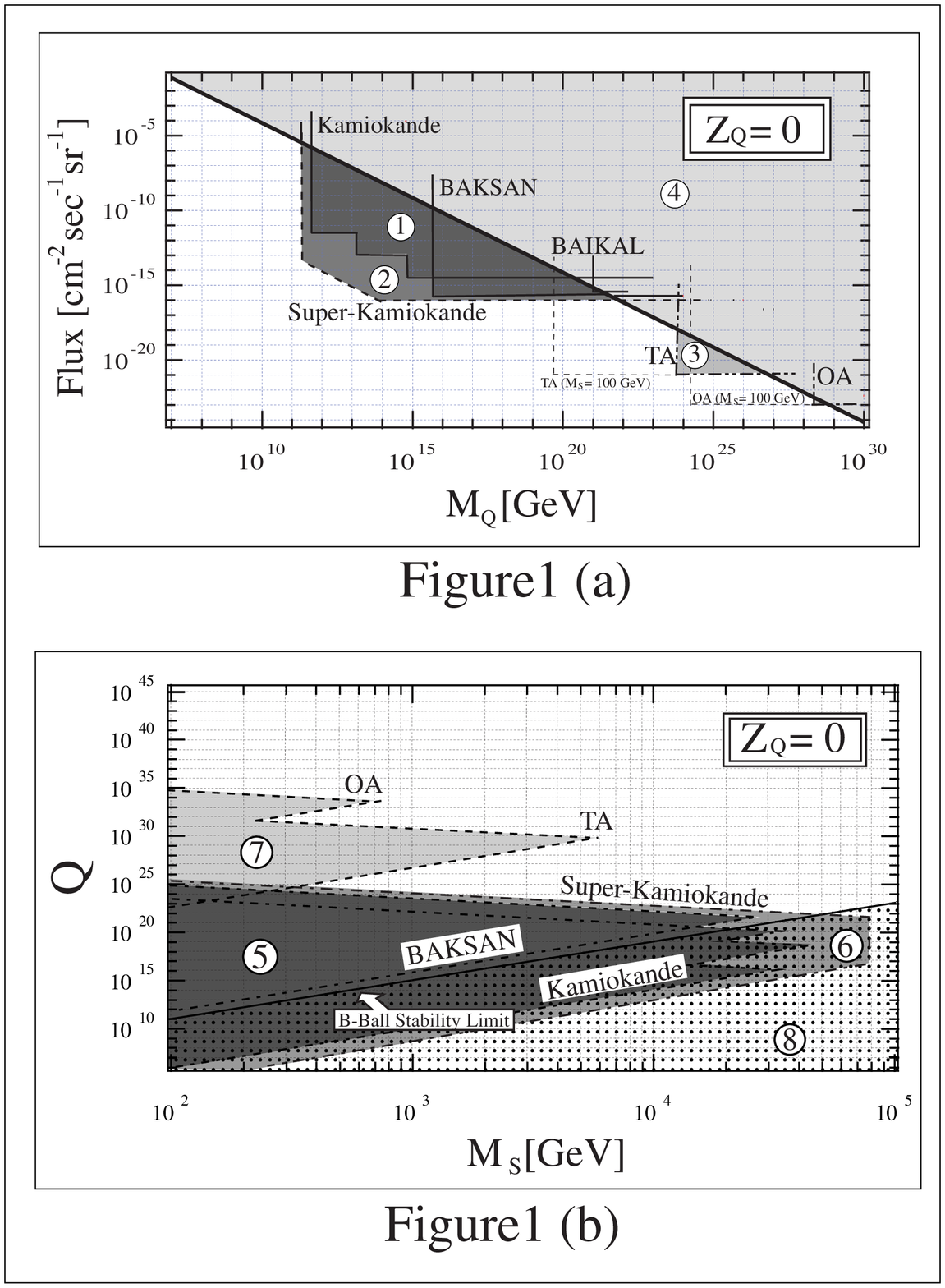}}
\caption{ Present limits on the electrically neural
dark-matter Q-balls from a paper by \textcite{arafune}.
}
\epsfclipoff
\label{fig_limits}
\end{figure}

Made of squarks, these Q-balls can interact strongly with ordinary matter,
{\em via} QCD.  However, if they are as heavy as the calculations show they
are, the strong interactions are not strong enough to force dark-matter
Q-balls to settle into the galactic disks.  Analyses of Q-ball formation
and partial evaporation allow one to relate the amounts of ordinary matter
and dark matter.  The observed ratio corresponds to Q-balls with baryon
number of about $10^{26\pm2}$, which is in agreement with the expected
Q-balls size from numerical simulations~\cite{kasuya1,kasuya3}, as well as
with the current experimental bounds summarized by~\textcite{arafune}, see
Fig.~\ref{fig_limits}.  A B-ball with baryon number $10^{26}$ is so heavy
that it could pass through ordinary stars with only a small change in its
velocity ($(\delta v/v) \sim 10^{-5}$).  Hence, despite the strong
interactions, B-balls make a good dark-matter candidate.

The unusually large mass for a dark matter
candidate means that the fluxes  are very small.
The experimental detection of this form of dark matter requires a large
detector size ({\em cf.} Fig.~\ref{fig_limits}).

Since B-balls have lower mass-to-baryon ratio than ordinary
nuclear matter, interactions of B-balls with
ordinary matter result in numerous ``proton decay'' events~\cite{kkst}.
Hence a Q-ball passing through a detector would produce a spectacular
signature.  The flux, however, is very low.  Hence, the strongest limits
come from the largest detectors, e.g. Super-Kamiokande (see
Fig.~\ref{fig_limits}).  Some astrophysical bounds have been
considered~\cite{starwreck}, but they do not yield very strong
constraints.
In addition
to the existing limits discussed in~\cite{arafune}, future experiments,
such as ANTARES, Ice Cube, {\em etc}, may be able to detect dark matter
B-balls, or rule
out the values of $Q$ that correspond to the correct amount of dark
matter.

We note that, although Q-balls are always present
in the spectrum of any SUSY extension of the Standard Model, their
production in the early universe requires the formation of an
Affleck--Dine condensate followed by its fragmentation.  Stable Q-balls are
too large to form in thermal plasma by accretion~\cite{gk,postma}.  In this
sense, an observation of stable dark-matter Q-balls would be evidence of
the Affleck--Dine process having taken place.

\subsubsection{Dark matter from unstable B-balls}
\label{dark_unstable}

If supersymmetry breaking is mediated by gravity, Q-balls are not stable as
they can decay into fermions.  However, Q-ball decay into fermions is a
slow process because the fermions quickly fill up the Fermi sea inside the
Q-ball, and further decays are limited by the rate of fermion
evaporation
through the surface.  The rate of Q-ball decay is, therefore, suppressed by
the surface-to-volume ratio~\cite{ccgm} as compared to that of free scalar
particles.  In a typical model, unstable baryonic Q-balls from the
Affleck--Dine condensate decay when the temperature is as low as a GeV.

The lightest supersymmetric particles (LSP) are among the decay products of
Q-balls.  B-balls can decay and produce dark matter in the form of
neutralinos at a time when they are out of
equilibrium~\cite{em_qb,em_d,fh_1}.  This presents another possibility for
producing dark matter from the AD condensate and relating its abundance to
that of ordinary matter.  The requirement that neutralinos not overclose
the universe constrains the parameter space of the MSSM~\cite{fh_1,fh_2}.

If the LSPs don't annihilate, the ratio of ordinary matter to dark matter
is simply~\cite{em_d}

\beq
\Omega_{\rm  matter }/
\Omega_{\rm  LSP } \sim f^{-1}
\left ( \frac{m_p}{m_\chi} \right )
\left ( \frac{n_{\rm B}}{n_\chi} \right )
\label{ratio_Omegas}
\eeq
where $f$ is the fraction of the condensate trapped in Q-balls
If $f\sim 10^{-3}$, this ratio is acceptable.

\begin{figure}
\setlength{\epsfxsize}{3in}
\centerline{\epsfbox{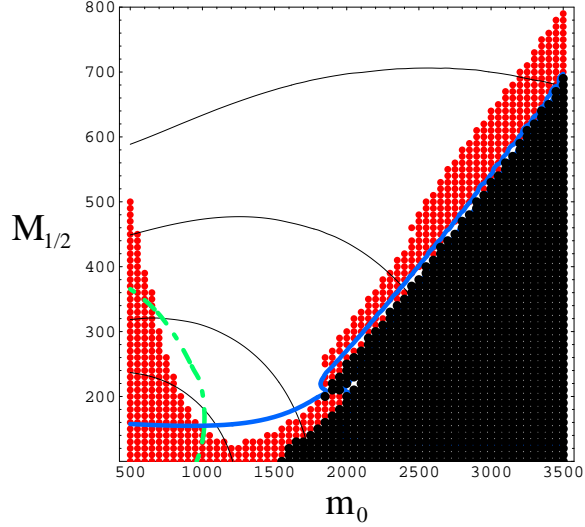}}
\caption{ The allowed range of parameters for non-thermal LSP dark matter
is very different from that in the standard freeze-out case.  The lightly
shaded region above the solid and the dashed lines is allowed for
non-thermal LSP dark matter in the minimal supergravity model with $\tan
\beta=40$~\cite{fh_1,fh_2}.  }
\label{figLSP}
\end{figure}

However, numerical simulations~\cite{kasuya1} and some analytical
calculations~\cite{f_McD} indicate that, in a wide
class of AD models, practically all the baryon number may be trapped in
Q-balls, that is $f\sim 1$.  If that's the case, the LSP would overclose
the universe, according to eq.~(\ref{ratio_Omegas}).  A solution, proposed
by~\textcite{fh_1,fh_2}, is to eliminate the unwanted overdensity of
neutralinos by using an LSP with a higher annihilation cross section.
The LSP in the MSSM is an admixture of several neutral fermions.  Depending
on the parameters in the mass matrix, determined largely by the soft SUSY
breaking terms, the LSP can be closely aligned with Bino (one of the
SUSY partners of the SU(2)$\times$U(1) gauge bosons), Higgsino (the
fermion counterpart of the Higgs boson), or with one of the other weak
eigenstates.  The traditional, freeze-out scenario for LSP production
favors the Bino-like LSP~\cite{kamionk}.  However, according
to~\textcite{fh_1,fh_2}, SUSY dark matter produced from the Affleck--Dine
process has to be in the form of a Higgsino-like LSP.
In this case, the ratio of matter densities is~\cite{fy_d}
\beq
\Omega_{\rm matter }/\Omega_{\rm LSP } =
10^{3-4} \left ( \frac{m_\phi^2}{\langle \sigma v \rangle ^{-1}_{\chi}
} \right ) \left ( \frac{m_p}{m_\chi} \right ) \delta_{\rm CP},
\eeq
where $\delta_{\rm CP} \sim 0.1$ is the effective CP violating phase of the
AD condensate. For a Higgsino-like LSP, which has $\langle \sigma v \rangle
\sim 10^{-(7-8)} {\rm GeV}^{-2}$, this yields an acceptable result.

This has important implications for both direct dark matter searches and
the collider searches for SUSY.  First, the parameter space of the MSSM
consistent with LSP dark matter is very different, depending on the
cosmological scenario at work, that is whether the LSPs froze out of
equilibrium~\cite{arnowitt} or were produced from the evaporation of AD
B-balls~\cite{fy_d}.  Second, higgsino and bino LSP's interact
differently with matter, so the sensitivity of direct dark-matter
searches also depends on the type of the LSP.

If supersymmetry is discovered, one will be able to determine the
properties of the LSP experimentally.  This will, in turn, provide some
information on the how the dark-matter SUSY particles could be produced.
The discovery of a Higgsino-like LSP would be a evidence in favor
of Affleck--Dine baryogenesis.  This is yet another way in which we might
be able to establish the true origin of matter-antimatter asymmetry.

\begin{figure}
\setlength{\epsfxsize}{4in}
\epsfclipon
\centerline{\epsfbox{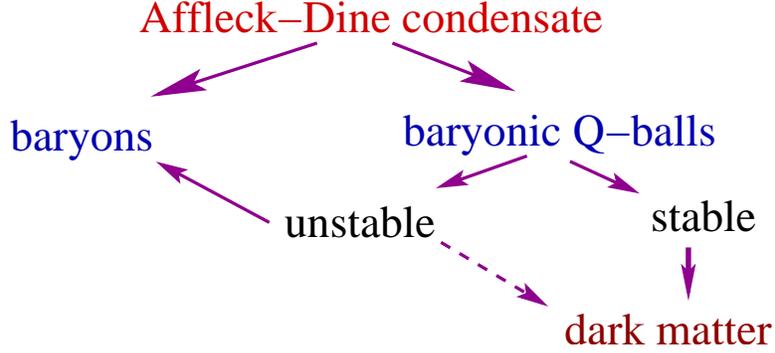}}
\caption{The fate of the AD condensate}
\epsfclipoff
\end{figure}

\section{Conclusions}

The origin of the matter-antimatter asymmetry is one of the great questions
in cosmology.  Yet we can obtain only limited information about the events
which gave rise to the baryon asymmetry by looking at the sky.  Filling out
the picture requires a deeper understanding of fundamental physical law.
One elegant possibility, that the Minimal Standard Model produced the
baryon number near the electroweak scale, is ruled out decisively by the
LEP bounds on the Higgs mass.  This is a bittersweet conclusion: while one
has to give up an elegant scenario, this is perhaps the strongest evidence
yet for physics beyond the Standard Model -- a precursor of future
discoveries.

Supersymmetry is widely regarded as a prime candidate for such new physics.
Theoretical arguments in favor of supersymmetry are based on the
naturalness of the scale hierarchy, the success of coupling unification
in supersymmetric theories, and the nearly ubiquitous role of
supersymmetry in string theory.
The upcoming LHC
experiments will put this hypothesis to a definitive test.  If low energy
supersymmetry exists, there are several ways in which it might play the
crucial role in baryogenesis.  It could conceivably revive the electroweak
baryogenesis scenario.  However, the phase transition in the MSSM is only
slightly stronger than that in the Standard Model; a noticeable improvement
forces one into a narrow corner of the MSSM parameter space, which may soon
be ruled out.

But supersymmetry opens a completely new and natural avenue for
baryogenesis.  If inflation took place in the early universe, for which we
have an increasing body of evidence, then formation of an Affleck--Dine
condensate and subsequent generation of some baryon asymmetry is natural.
In a wide class of models this process produces the observed baryon
asymmetry.  Perhaps more striking is that the process can lead
to very large baryon asymmetries.
This may be important in many cosmological
proposals where one produces substantial entropy at late times.

Finally, the same
process can produce dark matter, either in the form of stable SUSY Q-balls,
or in the form of a thermally or non-thermally produced LSP.
There are
hints that the relative closeness of matter and dark matter densities may
find its explanation in the same process as well.
If supersymmetry is discovered, given the success of inflation theory, the
Affleck--Dine scenario will appear quite plausible.

Other independent indications that the Affleck--Dine process took place in
the early universe may come from detection of dark matter.  One of the
great attractions of supersymmetry is that it can naturally
account for the dark matter.  The lightest
supersymmetric particle (LSP) is typically stable,
and is produced with an abundance in
a suitable range if the supersymmetry
breaking scale is of order 100's of GeV.
Its precise contribution to the energy
density of the universe depends on its annihilation cross section and mass.
A combination of accelerator limits and cosmology presently
allows for LSP dark matter in a range of parameters.  In this range,
the LSP, which is an admixture of several states, must be principally ``Bino''
(the supersymmetric partner of the $U(1)$ gauge boson)
if it is produced in the standard freeze-out scenario.  However, if future
detection will indicate that the LSP is Higgsino-like (i.e. primarily
the partner of  the Higgs boson), this kind of dark
matter could only arise from non-thermal production of the LSP from a
fragmented Affleck--Dine condensate~\cite{fh_1,fh_2,fy_d}.  Therefore,
although a standard Bino-like LSP is not inconsistent with the
Affleck--Dine scenario, a Higgsino-like LSP would provide a strong evidence
in its favor.  Likewise, a detection of stable baryonic Q-balls would be a
definitive confirmation that an Affleck--Dine condensate formed in the
early universe and fragmented into B-balls.  Since stable SUSY Q-balls must
be large, we know of no other cosmological scenario that could lead to
their formation.

Among other possibilities for baryogenesis, leptogenesis is also quite
plausible.  The discovery of neutrino mass, perhaps associated with a
rather low scale of new physics, certainly gives strong support to this
possibility.  The questions of what scales for this physics might be
compatible with inflation, and what implications this might have for the
underlying origin of neutrino mass are extremely important.  Some pieces of
the picture will be accessible to experiment, but many of the relevant
parameters, including the relevant CP violation, reside at a very high
scale.  Perhaps, in a compelling theory of neutrino flavor, some of these
questions can be pinned down.

Future experimental searches for supersymmetry, combined with the improving
cosmological data on CMBR and dark matter, will undoubtedly shed further
light on the origin of baryon asymmetry and will provide insight both
particle physics and cosmology.  The study of the baryon asymmetry has
already provided a compelling argument for new physics, and holds great
promise of new and exciting discoveries in the future.

\label{sec:concl}

\section{Acknowledgments}

This work was supported in part by the U.S. Department of Energy.
M.D. thanks for hospitality the Weizmann Institute, where he presented
lectures on which parts of this article are based.

\bibliographystyle{apsrmp}

\begin{thebibliography}{999}

\bibitem[Affleck and Dine, 1985]{ad}
I.~Affleck and M.~Dine,
Nucl.\ Phys.\ B {\bf 249}, 361 (1985).

\bibitem[Albrecht and Steinhardt, 1982]{steinhardt_infl}
A.~Albrecht and P.~J.~Steinhardt,
Phys.\ Rev.\ Lett.\  {\bf 48}, 1220 (1982).

\bibitem[Allahverdi,~Campbell,~and Ellis, 2000] {campbell_ellis}
R.~Allahverdi, B.~A.~Campbell and J.~R.~Ellis,
Nucl.\ Phys.\ B {\bf 579}, 355 (2000)
[arXiv:hep-ph/0001122].

\bibitem[Allahverdi {\em et al.}, 2002]{Allahverdi:2002vy}
R.~Allahverdi, A.~Mazumdar and A.~Ozpineci,
Phys.\ Rev.\ D {\bf 65}, 125003 (2002)
[arXiv:hep-ph/0203062].

\bibitem[Altarelli,~Feruglio and Masina, 2000]{altarelli}
G.~Altarelli, F.~Feruglio and I.~Masina,
JHEP {\bf 0011}, 040 (2000)
[arXiv:hep-ph/0007254].

\bibitem[Anisimov and Dine, 2000]{anisimov_dine}
A. Anisimov and M, Dine,
Nucl.Phys.B619:729-740,2001, hep-ph/0008058

\bibitem[Arafune {\em et al.}, 2000]{arafune}
J.~Arafune, T.~Yoshida, S.~Nakamura and K.~Ogure,
Phys.\ Rev.\ D {\bf 62}, 105013 (2000)
[arXiv:hep-ph/0005103].

\bibitem[Arnold and Espinosa, 1993]{twoloop_2}
P.~Arnold and O.~Espinosa,
Phys.\ Rev.\ D {\bf 47}, 3546 (1993)
[Erratum-ibid.\ D {\bf 50}, 6662 (1994)].

\bibitem[Arnold and McLerran, 1988]{arnoldmclerran}
P.~Arnold, and L.~Mclerran, Phys.Rev {\bf  D37} (1988) 1020.

\bibitem[Arnold,~Son,~and Yaffe, 1997]{asy}
P.~Arnold, D.~Son and L.~G.~Yaffe,
Phys.\ Rev.\ D {\bf 55}, 6264 (1997)
[arXiv:hep-ph/9609481].
Phys.\ Rev.\ D {\bf 62}, 023512 (2000)
[arXiv:hep-ph/0002285].


\bibitem[Arnowitt and Dutta, 2002]{arnowitt}
R.~Arnowitt and B.~Dutta,
arXiv:hep-ph/0211042.

\bibitem[Babu,~Pati and Wilczek, 2000]{wilczek_gut}
K.~S.~Babu, J.~C.~Pati and F.~Wilczek,
Nucl.\ Phys.\ B {\bf 566}, 33 (2000).

\bibitem[Bajc,~Perez and Senjanovic, 2002]{senjanovic}
B.~Bajc, P.~F.~Perez and G.~Senjanovic,
Phys.\ Rev.\ D {\bf 66}, 075005 (2002).

\bibitem[Bagnasco and Dine, 1993] {twoloop_1}
J.~E.~Bagnasco and M.~Dine,
Phys.\ Lett.\ B {\bf 303}, 308 (1993).


\bibitem[Banerjee and Jedamzik, 2000]{jedamzik}
R.~Banerjee and K.~Jedamzik,
Phys.\ Lett.\ B {\bf 484}, 278 (2000)
[arXiv:hep-ph/0005031].

\bibitem[Banks {\em et al.}, 1988]{banksglobal}
T. Banks, L.J. Dixon, D. Friedan and E. Martinec,
Nucl.Phys. {\bf B299}, 613 (1988).

\bibitem[Barr,~Segr\`e and Weldon, 1979]{barr}
S.~M.~Barr, G.~Segr\`e and H.~A.~Weldon,
Phys.\ Rev.\ D {\bf 20}, 2494 (1979).

\bibitem[Belotsky {\em et al.}, 1998]{anti_3}
K.~M.~Belotsky, Y.~A.~Golubkov, M.~Y.~Khlopov, R.~V.~Konoplich and
A.~S.~Sakharov,
arXiv:astro-ph/9807027.


\bibitem[Bennett {\em et~al.}, 2003]{map}
C.L.~Bennett {\em et al.} [WMAP collaboration], astro-ph/0302207.

\bibitem[Bodeker {\em et al.}, 1997]{loops2}
D.~Bodeker, P.~John, M.~Laine and M.~G.~Schmidt,
Nucl.\ Phys.\ B {\bf 497}, 387 (1997).

\bibitem[Burles,~Nollett,~Turner, 2001]{burles}
S.~Burles, K.~M.~Nollett and M.~S.~Turner,
Phys.\ Rev.\ D {\bf 63}, 063512 (2001);
Astrophys.\ J.\  {\bf 552}, L1 (2001).

\bibitem[Callan and Coleman, 1977]{tunn2} C.G.~Callan and S.~Coleman,
Phys. Rev. {\bf D16}, 1762 (1977).

\bibitem[Carena,~Quiros,~and Wagner, 1998]{carena}
M.~Carena, M.~Quiros and C.~E.~Wagner,
Phys.\ Lett.\ B {\bf 380}, 81 (1996); Nucl.\ Phys.\ B {\bf 524}, 3 (1998).

\bibitem[Carena {\em et al.}, 2001]{Carena:2000id}
M.~Carena, J.~M.~Moreno, M.~Quiros, M.~Seco and C.~E.~Wagner,
Nucl.\ Phys.\ B {\bf 599}, 158 (2001)
[arXiv:hep-ph/0011055].

\bibitem[Carena,~Quiros,~Seco,~and Wagner, 2002]{carena_2002}
M.~Carena, M.~Quiros, M.~Seco and C.~E.~Wagner,
Nucl.\ Phys.\ B {\bf 650}, 24 (2003).

\bibitem[Chang,~Chang and Keung, 2002]{mssm_constraint}
D.~Chang, W.~F.~Chang and W.~Y.~Keung,
arXiv:hep-ph/0205084.

\bibitem[Cline, 2002]{cline_review}
J.~M.~Cline,
Pramana {\bf 54}, 1 (2000)
[Pramana {\bf 55}, 33 (2000)];
arXiv:hep-ph/0201286.

\bibitem[Cline,~Joyce and Kainulainen, 1998]{cjk}
J.~M.~Cline, M.~Joyce and K.~Kainulainen,
Phys.\ Lett.\ B {\bf 417}, 79 (1998); {\it ibid.},\ B {\bf 448}, 321 (1999);
JHEP {\bf 0007}, 018 (2000).


\bibitem[Cline {\em et al.}, 1996]{cline-kimmo}
J.~M.~Cline, K.~Kainulainen and A.~P.~Vischer,
Phys.\ Rev.\ D {\bf 54}, 2451 (1996).

\bibitem[Cline and Moore, 1998]{cline-moore}
J.~M.~Cline and G.~D.~Moore,
Phys.\ Rev.\ Lett.\  {\bf 81}, 3315 (1998).

\bibitem[Cohen,~Kaplan,~and Nelson, 1991]{ckn_spont}
A.~G.~Cohen, D.~B.~Kaplan,~and A.~E.~Nelson,
Phys.\ Lett.\ B {\bf 263}, 86 (1991).

\bibitem[Cohen,~Kaplan,~and Nelson, 1992]{ckn_susy}
A.~G.~Cohen,D.~B.~Kaplan, and A.~E.~Nelson,
Phys.\ Lett.\ B {\bf 297}, 111 (1992).

\bibitem[Cohen,~Kaplan and Nelson, 1993]{ckn_review}
A.~G.~Cohen, D.~B.~Kaplan and A.~E.~Nelson,
Ann.\ Rev.\ Nucl.\ Part.\ Sci.\  {\bf 43}, 27 (1993)
[arXiv:hep-ph/9302210].

\bibitem[Cohen,~Coleman,~Georgi,~and Manohar, 1986]{ccgm}
A.~G.~Cohen, S.~R.~Coleman, H.~Georgi and A.~Manohar,
Nucl.\ Phys.\ B {\bf 272}, 301 (1986).

\bibitem[Cohen,~De Rujula,~and~Glashow, 1998]{cohenetal}
A.~G.~Cohen, A.~De Rujula and S.~L.~Glashow,
Astrophys.\ J.\  {\bf 495}, 539 (1998)


\bibitem[Coleman, 1977] {tunn1} S.~Coleman, Phys. Rev. {\bf D15}, 2929
  (1977).

\bibitem[Coleman, 1985] {coleman}
S.~Coleman, Nucl. Phys. {\bf B262} (1985) 263.

\bibitem[Coleman, 1989] {colemanuses}
S. ~Coleman, {\it Aspects of Symmetry}, Cambridge University Press,
Cambridge, 1989.

\bibitem[Coleman {\em et al.}, 1978] {cgm}
S.~Coleman, V.~Glaser and A.~Martin, Comm. Math. Phys. {\bf
58} (1978) 211.

\bibitem[Copeland {\em et  al.}, 2001]{copeland}
E.~J.~Copeland, D.~Lyth, A.~Rajantie and M.~Trodden,
Phys.\ Rev.\ D {\bf 64}, 043506 (2001)
[arXiv:hep-ph/0103231].

\bibitem[Cornwall and Kusenko, 2000]{ck}
J.~M.~Cornwall and A.~Kusenko,
Phys.\ Rev.\ D {\bf 61}, 103510 (2000)

\bibitem[Cornwall {\em et al.}, 2001]{cgk}
J.~M.~Cornwall, D.~Grigoriev and A.~Kusenko,
Phys.\ Rev.\ D {\bf 64}, 123518 (2001)

\bibitem[Correia and Schmidt, 2001]{PaccettiCorreia:2001uh}
F.~Paccetti Correia and M.~G.~Schmidt,
Eur.\ Phys.\ J.\ C {\bf 21}, 181 (2001)
[arXiv:hep-th/0103189].

\bibitem[Coughlan {\em et al.}, 1983]{moduli}
G.~D.~Coughlan, W.~Fischler, E.~W.~Kolb, S.~Raby and G.~G.~Ross,
Phys.\ Lett.\ B {\bf 131}, 59 (1983).

\bibitem[Csikor {\em et al.}, 2000]{pt_mssm1}
F.~Csikor, Z.~Fodor, P.~Hegedus, A.~Jakovac, S.~D.~Katz and A.~Piroth,
Phys.\ Rev.\ Lett.\  {\bf 85}, 932 (2000).

\bibitem[Csikor {\em et al.}, 1999]{EW_l1}
F.~Csikor, Z.~Fodor and J.~Heitger,
Phys.\ Rev.\ Lett.\  {\bf 82}, 21 (1999)
[arXiv:hep-ph/9809291].

\bibitem[Davies,~Froggatt,~and Moorhouse, 1996]{nmssm1}
A.~T.~Davies, C.~D.~Froggatt and R.~G.~Moorhouse,
Phys.\ Lett.\ B {\bf 372}, 88 (1996).

\bibitem[de Bernardis {\it et al.}, 2000]{CMBR_BOOMERANG_1}
P.~de Bernardis {\it et al.} [Boomerang Collaboration],
Nature {\bf 404}, 955 (2000).

\bibitem[Dermisek,Mafi and Raby, 2001]{raby}
R.~Dermisek, A.~Mafi and S.~Raby,
Phys.\ Rev.\ D {\bf 63}, 035001 (2001)
[arXiv:hep-ph/0007213].

\bibitem[Dine, 1996]{brush}
M.~Dine,
arXiv:hep-ph/9612389.

\bibitem[Dine, 2002]{dine_moduli}
M.~Dine,
Phys.\ Lett.\ B {\bf 482}, 213 (2000)
[arXiv:hep-th/0002047].

\bibitem[Dine {\em et.al}, 1990]{dinetal}
M.~Dine, O.~Lechtenfeld, B.~Sakita, W.~Fischler and J.~Polchinski,
Nucl.\ Phys.\ B {\bf 342}, 381 (1990).

\bibitem[Dine  {\em et.al}, 1991]{dine_mssm1}
M.~Dine, P.~Huet, R.~J.~Singleton and L.~Susskind,
Phys.\ Lett.\ B {\bf 257}, 351 (1991);

\bibitem[Dine,~Huet,~and Singleton, 1992]{dine_mssm2}
M.~Dine, P.~Huet and R.~J.~Singleton,
Nucl.\ Phys.\ B {\bf 375}, 625 (1992).

\bibitem[Dine,~Randall,~and Thomas, 1995]{drt_SUSYbreaking}
M.~Dine, L.~Randall and S.~Thomas,
Phys.\ Rev.\ Lett.\  {\bf 75}, 398 (1995)
[arXiv:hep-ph/9503303].

\bibitem[Dine,~Randall,~and Thomas, 1996]{drt}
M.~Dine, L.~Randall and S.~Thomas,
Nucl.\ Phys.\ B {\bf 458}, 291 (1996)
[arXiv:hep-ph/9507453].

\bibitem[Dine and Thomas, 1994]{dt}
M.~Dine and S.~Thomas,
Phys.\ Lett.\ B {\bf 328}, 73 (1994)
[arXiv:hep-ph/9401265].

\bibitem[Dolgov and Silk, 1993]{anti_1}
A.~Dolgov and J.~Silk,
Phys.\ Rev.\ D {\bf 47}, 4244 (1993).

\bibitem[Dvali, 1995]{dvali_SUSYbreaking}
G.~Dvali,
Phys.\ Lett.\ B {\bf 355}, 78 (1995)
[arXiv:hep-ph/9503375].

\bibitem[Dvali, Kusenko, and Shaposhnikov, 1998] {dks}
G.~Dvali, A.~Kusenko, M.~Shaposhnikov,
Phys.\ Lett.\  B {\bf 417}, 99 (1998)

\bibitem[Ellis {\it et al.}, 1992 ]{eglns}
J.~R.~Ellis, G.~B.~Gelmini, J.~L.~Lopez, D.~V.~Nanopoulos and S.~Sarkar,
Nucl.\ Phys.\ B {\bf 373}, 399 (1992).

\bibitem[KamLAND, 2002]{kamland}
K.~Eguchi {\it et al.}  [KamLAND Collaboration],
Phys.\ Rev.\ Lett.\  {\bf 90}, 021802 (2003)
[arXiv:hep-ex/0212021].

\bibitem[Enqvist {\em et al.}, 2000]{Enqvist:2000gq}
K.~Enqvist, A.~Jokinen and J.~McDonald,
Phys.\ Lett.\ B {\bf 483}, 191 (2000)
[arXiv:hep-ph/0004050].

\bibitem[Enqvist {\em et al.}, 2001] {enqvist_num}
K.~Enqvist, A.~Jokinen, T.~Multamaki and I.~Vilja,
Phys.\ Rev.\ D {\bf 63}, 083501 (2001)
[arXiv:hep-ph/0011134].

\bibitem[Enqvist and Mazumdar, 2002] {Enqvist:2003gh}
K.~Enqvist and A.~Mazumdar,
arXiv:hep-ph/0209244.

\bibitem[Enqvist and McDonald, 1998]{em_qb}
K.~Enqvist and J.~McDonald,
Phys.\ Lett.\ B {\bf 425}, 309 (1998);
Phys.\ Lett.\ B {\bf 440}, 59 (1998).

\bibitem[Enqvist and McDonald, 1999]{em_d}
K.~Enqvist and J.~McDonald,
Nucl.\ Phys.\ B {\bf 538}, 321 (1999)
[arXiv:hep-ph/9803380].

\bibitem[Espinosa,~Quir{\'o}s,~and Zwirner, 1993]{eqz}
J.R.~Espinosa, M.~Quir{\'o}s and F.~Zwirner,
Phys.\ Lett.\ B {\bf 307}, 106 (1993).

\bibitem[Espinosa, 1996]{loops1}
J.~R.~Espinosa,
Nucl.\ Phys.\ B {\bf 475}, 273 (1996). 

\bibitem[Farakos {\em et al.}, 1994]{twoloop_3}
K.~Farakos, K.~Kajantie, K.~Rummukainen and M.~E.~Shaposhnikov,
Nucl.\ Phys.\ B {\bf 425}, 67 (1994).

\bibitem[Farakos {\em et al.}, 1995]{loops_and_lattice}
K.~Farakos, K.~Kajantie, K.~Rummukainen and M.~E.~Shaposhnikov,
Nucl.\ Phys.\ B {\bf 442}, 317 (1995).

\bibitem[Felder {\em et al.}, 2001]{tachionic_preheating}
G.~N.~Felder, J.~Garc\'ia-Bellido, P.~B.~Greene, L.~Kofman, A.~D.~Linde and
I.~Tkachev,
Phys.\ Rev.\ Lett.\  {\bf 87}, 011601 (2001)
[arXiv:hep-ph/0012142].

\bibitem[Friedmann and Witten, 2002]{witten_GUT}
T.~Friedmann and E.~Witten,
arXiv:hep-th/0211269.

\bibitem[Frieman, Olilnto, Gleiser, and Alcock, 1989]{frieman}
J.~A.~Frieman, A.~V.~Olinto, M.~Gleiser, C.~Alcock:
Phys.\ Rev.\  D {\bf 40}, 3241 (1989)

\bibitem[Super-K, 1998]{superK_atmos}
Y.~Fukuda {\it et al.}  [Super-Kamiokande Collaboration],
Phys.\ Rev.\ Lett.\  {\bf 81}, 1562 (1998)
[arXiv:hep-ex/9807003].

\bibitem[Super-K, 2001]{superK_solar}
S.~Fukuda {\it et al.}  [Super-Kamiokande Collaboration],
Phys.\ Rev.\ Lett.\  {\bf 86}, 5656 (2001)
[arXiv:hep-ex/0103033].

\bibitem[Fujii and Hamaguchi, 2002a]{fh_1}
M.~Fujii and K.~Hamaguchi,
Phys.\ Lett.\ B {\bf 525}, 143 (2002)
.

\bibitem[Fujii and Hamaguchi, 2002b]{fh_2}
M.~Fujii and K.~Hamaguchi,
Phys.\ Rev.\ D {\bf 66}, 083501 (2002)


\bibitem[Fujii and Yanagida, 2002a]{fy_d}
M.~Fujii and T.~Yanagida,
Phys.\ Lett.\ B {\bf 542}, 80 (2002)
[arXiv:hep-ph/0206066].

\bibitem[Fujii amd Yanagida, 2002b]{fy_d2}
M.~Fujii and T.~Yanagida,
arXiv:hep-ph/0207339.


\bibitem[Fukugita and Yanagida, 1986]{fy}
M.~Fukugita and T.~Yanagida,
Phys.\ Lett.\ B {\bf 174}, 45 (1986).


\bibitem[Friedberg,~Lee,~and Sirlin, 1976] {fls}
R.~Friedberg, T.~D.~Lee,
A.~Sirlin: Phys. Rev. D {\bf 13}, 2739 (1976);

\bibitem[Garc\'ia-Bellido {\em et al.}, 1999]{ggks}
J.~Garc\'ia-Bellido, D.~Grigoriev, A.~Kusenko, and
M.~Shaposhnikov, Phys. Rev. D {\bf 60}, 123504 (1999).

\bibitem[Garc\'ia-Bellido and Morales, 2001]{Garcia-Bellido:2001cb}
J.~Garc\'ia-Bellido and E.~Ruiz Morales,
Phys.\ Lett.\ B {\bf 536}, 193 (2002)
[arXiv:hep-ph/0109230].

\bibitem[Garc\'ia-Bellido {\em et al.}, 2003]{Garcia-Bellido:2003}
J.~Garc\'ia-Bellido, M.~Garcia-Perez and A.~Gonzalez-Arroyo,
arXiv:hep-ph/0304285.


\bibitem[Georgi and Glashow, 1974]{gg}
H.~Georgi and S.~L.~Glashow,
Phys.\ Rev.\ Lett.\  {\bf 32}, 438 (1974).

\bibitem[German,~Ross,~and Sarkar, 2001]{grs}
G.~German, G.~Ross and S.~Sarkar,
Nucl.\ Phys.\ B {\bf 608}, 423 (2001)
[arXiv:hep-ph/0103243].


\bibitem[Gherghetta,~Kolda and Martin, 1996] {flat}
T.~Gherghetta, C.~F.~Kolda and S.~P.~Martin,
Nucl.\ Phys.\ B {\bf 468}, 37 (1996).

\bibitem[Gonzalez-Garcia {\em et al.}, 2001]{nirCP}
M.~C.~Gonzalez-Garcia, Y.~Grossman, A.~Gusso and Y.~Nir,
Phys.\ Rev.\ D {\bf 64}, 096006 (2001)
[arXiv:hep-ph/0105159].


\bibitem[Gonzalez-Garcia and Nir, 2002]{nirreview}
M.~C.~Gonzalez-Garcia and Y.~Nir,
arXiv:hep-ph/0202058.

\bibitem[Griest and Kolb, 1989]{gk}
K.~Griest, E.~W.~Kolb:
Phys.\ Rev.\  D {\bf 40}, 3231 (1989);

\bibitem[Grigoriev,Rubakov and Shaposhnikov, 1989a]{Grigoriev}
D.~Y.~Grigoriev, V.~A.~Rubakov and M.~E.~Shaposhnikov,
Phys.\ Lett.\ B {\bf 216}, 172 (1989a).

\bibitem[Grigoriev,Rubakov and Shaposhnikov, 1989b]{Grigoriev:ub}
D.~Y.~Grigoriev, V.~A.~Rubakov and M.~E.~Shaposhnikov,
Nucl.\ Phys.\ B {\bf 326}, 737 (1989b).

\bibitem[Gurtler {\em et al.}, 1997]{EW_l2}
M.~Gurtler, E.~M.~Ilgenfritz and A.~Schiller,
Phys.\ Rev.\ D {\bf 56}, 3888 (1997)
[arXiv:hep-lat/9704013].

\bibitem[Guth, 1981]{guth}
A.~H.~Guth,
Phys.\ Rev.\ D {\bf 23}, 347 (1981).

\bibitem[Harvey and Turner, 1990]{Harvey:1990qw}
J.~A.~Harvey and M.~S.~Turner,
Phys.\ Rev.\ D {\bf 42}, 3344 (1990).

\bibitem[Hisano et al., 2001]{Hisano:2001dr}
J.~Hisano, M.~M.~Nojiri and N.~Okada,
Phys.\ Rev.\ D {\bf 64}, 023511 (2001)
[arXiv:hep-ph/0102045].

\bibitem[Particle Data Group, 2002] {pdg}
K. Hagiwara {\em et al.}, Phys. Rev. D66, 010001 (2002)

\bibitem[S.~Hanany {\it et al.}, 2000]{CMBR_MAXIMA}
S.~Hanany {\it et al.},
Astrophys.\ J.\ {\bf 545}, L5 (2000).

\bibitem[Huber {\it et al.}, 2001]{nmssm4}
S.~J.~Huber, P.~John and M.~G.~Schmidt,
Eur.\ Phys.\ J.\ C {\bf 20}, 695 (2001).

\bibitem[Huber and Schmidt, 2001]{nmssm2}
S.~J.~Huber and M.~G.~Schmidt,
Nucl.\ Phys.\ B {\bf 606}, 183 (2001).

\bibitem[Huet and Nelson, 1996]{huet_nelson}
P.~Huet and A.~E.~Nelson,
Phys.\ Rev.\ D {\bf 53}, 4578 (1996)
[arXiv:hep-ph/9506477].

\bibitem[Jungman,~Kamionkowski,~and Griest, 1996] {kamionk}
G.~Jungman, M.~Kamionkowski and K.~Griest,
Phys.\ Rept.\  {\bf 267}, 195 (1996)
[arXiv:hep-ph/9506380].

\bibitem[Kainulainen {\em et al.}, 2001]{nmssm3}
K.~Kainulainen, T.~Prokopec, M.~G.~Schmidt and S.~Weinstock,
JHEP {\bf 0106}, 031 (2001).


\bibitem[Kajantie {\em et al.}, 1996] {EW_num1}
K.~Kajantie, M.~Laine, K.~Rummukainen and M.~E.~Shaposhnikov,
Nucl.\ Phys.\ B {\bf 466}, 189 (1996);
Phys.\ Rev.\ Lett.\  {\bf 77}, 2887 (1996).

\bibitem[Kajantie {\em et al.}, 1997]{EW_num2}
K.~Kajantie, M.~Laine, K.~Rummukainen and M.~E.~Shaposhnikov,
Nucl.\ Phys.\ B {\bf 493}, 413 (1997) .

\bibitem[Kajantie {\em et al.}, 1999] {EW_num_B}
K.~Kajantie, M.~Laine, J.~Peisa, K.~Rummukainen and M.~E.~Shaposhnikov,
Nucl.\ Phys.\ B {\bf 544}, 357 (1999).

\bibitem[Kallosh {\em et al.}, 2000]{kallosh}
R.~Kallosh, L.~Kofman, A.~D.~Linde and A.~Van Proeyen,
Phys.\ Rev.\ D {\bf 61}, 103503 (2000).

\bibitem[Kaplan, 1992]{kaplan}
D.~B.~KS.~Kasuya and M.~Kawasaki,
Phys.\ Rev.\ D {\bf 61}, 041301 (2000).

\bibitem[Karsch {\em et al.}, 1997]{EW_l3}
F.~Karsch, T.~Neuhaus, A.~Patkos and J.~Rank,
Nucl.\ Phys.\ Proc.\ Suppl.\  {\bf 53}, 623 (1997)
[arXiv:hep-lat/9608087].

\bibitem[Kasuya and Kawasaki, 2000a]{kasuya1}
S.~Kasuya and M.~Kawasaki,
Phys.\ Rev.\ D {\bf 61}, 041301 (2000).

\bibitem[Kasuya and Kawasaki, 2000b]{kasuya2}
S.~Kasuya and M.~Kawasaki,
Phys.\ Rev.\ D {\bf 62}, 023512 (2000)
[arXiv:hep-ph/0002285].


\bibitem[Kasuya and Kawasaki, 2001]{kasuya3}
S.~Kasuya and M.~Kawasaki,
Phys.\ Rev.\ D {\bf 64}, 123515 (2001). 


\bibitem[Kasuya {\em et al.}, 2002]{kasuya_moduli}
S.~Kasuya, M.~Kawasaki and F.~Takahashi,
Phys.\ Rev.\ D {\bf 65}, 063509 (2002)
[arXiv:hep-ph/0108171].

\bibitem[Khlebnikov and Shaposhnikov, 1988]{Khlebnikov}
S.~Y.~Khlebnikov and M.~E.~Shaposhnikov,
Nucl.\ Phys.\ B {\bf 308}, 885 (1988).

\bibitem[Khlopov {\em et al.}, 2000]{anti_2}
M.~Y.~Khlopov, S.~G.~Rubin and A.~S.~Sakharov,
Phys.\ Rev.\ D {\bf 62}, 083505 (2000)
[arXiv:hep-ph/0003285].

\bibitem[Klinkhammer and Manton, 1984]{manton1}
F. Klinkhammer and N. Manton, Phys. Rev. {\bf D30} (1984) 2212.

\bibitem[Kobayashi and Maskawa, 1973]{ckm}
M.~Kobayashi and T.~Maskawa,
Prog.\ Theor.\ Phys.\  {\bf 49}, 652 (1973).

\bibitem[Kofman,~Linde and Starobinsky, 1996]{kls}
L.~Kofman, A.~Linde, and A.~A.~Starobinsky,
Phys. Rev. Lett. {\bf 73}, 3195 (1994), {\it ibid.} {\bf 76}, 1011 (1996);
Phys. Rev. D {\bf 56}, 3258 (1997).

\bibitem[Kolb and Turner, 1990]{kolb}
E.W. Kolb and M.S. Turner, {\it The Early Universe},
Adddison-Wesley, Reading, MA (1990).

\bibitem[Kolda and March-Russell, 1999]{Kolda:1998kc}
C.~F.~Kolda and J.~March-Russell,
Phys.\ Rev.\ D {\bf 60}, 023504 (1999)
[arXiv:hep-ph/9802358].

\bibitem[Krauss and Trodden, 1999] {kt}
L.~M.~Krauss and M.~Trodden, Phys. Rev. Lett.{\bf 83}, 1502 (1999).

\bibitem[Kusenko, 1997a] {ak_qb} A.~Kusenko:
Phys.\ Lett.\  B {\bf 404}, 285 (1997), hep-th/9704073

\bibitem[Kusenko, 1997b]{ak_mssm} A.~Kusenko:
Phys.\ Lett.\  B {\bf 405}, 108 (1997)


\bibitem[Kusenko and Shaposhnikov, 1998] {ks}
A.~Kusenko, M.~Shaposhnikov:
Phys.\ Lett.\  B {\bf 418}, 46 (1998)

\bibitem[Kusenko {\em et al.}, 1998a]{kkst}
A.~Kusenko, V.~Kuzmin, M.~Shaposhnikov, P.~G.~Tinyakov:
Phys.\ Rev.\ Lett.\  {\bf 80}, 3185 (1998)

\bibitem[Kusenko,~Shaposhnikov,~and~Tinyakov, 1998]{kst}
A.~Kusenko, M.~Shaposhnikov, P.~G.~Tinyakov:
Pisma Zh.\ Eksp.\ Teor.\ Fiz.\  {\bf 67}, 229 (1998)

\bibitem[Kusenko {\em et al.}, 1998c]{starwreck}
A.~Kusenko, M.~E.~Shaposhnikov, P.~G.~Tinyakov and I.~I.~Tkachev,
Phys.\ Lett.\ B {\bf 423}, 104 (1998)
[arXiv:hep-ph/9801212].

\bibitem[Kuzmin,~Rubakov,~and Shaposhnikov, 1985]{krs}
V.~A.~Kuzmin, V.~A.~Rubakov and M.~E.~Shaposhnikov,
Phys.\ Lett.\ B {\bf 155}, 36 (1985).

\bibitem[Laine, 1994]{twoloop_4}
M.~Laine,
Phys.\ Lett.\ B {\bf 335}, 173 (1994).


\bibitem[Laine and Rummukainen, 1998]{pt_mssm}
M.~Laine and K.~Rummukainen,
Nucl.\ Phys.\ B {\bf 535}, 423 (1998)
[arXiv:hep-lat/9804019].

\bibitem[Laine and Shaposhnikov, 1998]{ls}
M.~Laine and M.~E.~Shaposhnikov,
Nucl.\ Phys.\ B {\bf 532}, 376 (1998)
[arXiv:hep-ph/9804237].

\bibitem[Lee {\em et al.}, 1989]{gauged_Q}
K.~M.~Lee, J.~A.~Stein-Schabes, R.~Watkins and L.~M.~Widrow,
Phys.\ Rev.\ D {\bf 39}, 1665 (1989).

\bibitem[Lee, 1974]{lw1} T.~D.~Lee, Phys. Rev. D {\bf 8}, 1226 (1973);
Phys. Reports {\bf 9}, 143 (1974);

\bibitem[Lee and Pang, 1992]{tdlee}
T.~D.~Lee and Y.~Pang,
Phys.\ Rept.\  {\bf 221}, 251 (1992).

\bibitem[Linde, 1982]{linde_infl}
A.~D.~Linde,
Phys.\ Lett.\ B {\bf 108}, 389 (1982).

\bibitem[Linde, 1990]{lindebook}
A. Linde, {\it Particle Physics and Inflationary Cosmology},
Harwood, Chur, Switzerland (1990).

\bibitem[Lyth, 1999]{lyth} 
D.~H.~Lyth,
Phys.\ Lett.\ B {\bf 466}, 85 (1999)
[arXiv:hep-ph/9908219].

\bibitem[Manton, 1983]{manton}
N. Manton, Phys. Rev. {\bf D28} (1983) 2019.

\bibitem[Mazumdar {\em et al.}, 2002]  {Mazumdar:2002rn}
A.~Mazumdar, K.~Enqvist and S.~Kasuya,
arXiv:hep-ph/0210241.

\bibitem[McDonald, 2001]{f_McD}
J.~McDonald,
JHEP {\bf 0103}, 022 (2001)
[arXiv:hep-ph/0012369].

\bibitem[Multamaki, 2001]{Multamaki:2001az}
T.~Multamaki,
Phys.\ Lett.\ B {\bf 511}, 92 (2001)
[arXiv:hep-ph/0102339].

\bibitem[Multamaki and I.~Vilja, 2000]{Multamaki:1999an}
T.~Multamaki and I.~Vilja,
Nucl.\ Phys.\ B {\bf 574}, 130 (2000)
[arXiv:hep-ph/9908446].

\bibitem[Multamaki and Vilja, 2002]{Multamaki:2002hv}
T.~Multamaki and I.~Vilja,
Phys.\ Lett.\ B {\bf 535}, 170 (2002)
[arXiv:hep-ph/0203195].

\bibitem[Murayama and Pierce, 2002]{murayama}
H.~Murayama, and A.~Pierce,
Phys.Rev. {\bf D65}, 055009 (2002).

\bibitem[Murayama {\em et al.}, 1993]{murayama1}
H.~Murayama, H.~Suzuki, T.~Yanagida and J.~Yokoyama,
Phys.\ Rev.\ Lett.\  {\bf 70}, 1912 (1993).

\bibitem[Murayama and Yanagida, 1994]{murayama2}
H.~Murayama and T.~Yanagida,
Phys.\ Lett.\ B {\bf 322}, 349 (1994)
[arXiv:hep-ph/9310297].

\bibitem[Netterfield {\it et al.}, 2002]{CMBR_BOOMERANG_2}
C.~B.~Netterfield {\it et al.}  [Boomerang Collaboration],
Astrophys.\ J.\  {\bf 571}, 604 (2002).

\bibitem[Netterfield {\it et al.}, 2002]{CMBR_DASI}
C.~Pryke, N.~W.~Halverson, E.~M.~Leitch, J.~Kovac, J.~E.~Carlstrom,
W.~L.~Holzapfel and M.~Dragovan,
Astrophys.\ J.\  {\bf 568}, 46 (2002).


\bibitem[Kirkman {\em et al.}, 2003]{kirkman}
D.~Kirkman, D.~Tytler, N.~Suzuki, J.~M.~O'Meara and D.~Lubin,
arXiv:astro-ph/0302006.


\bibitem[Pascoli {\em et al.}, 2003]{lepto_CP}
S.~Pascoli, S.~T.~Petcov and W.~Rodejohann,
arXiv:hep-ph/0302054.

\bibitem[Pilaftsis, 2002] {mssm_constraint1}
A.~Pilaftsis,
Nucl.\ Phys.\ B {\bf 644}, 263 (2002).

\bibitem[Postma, 2002]{postma}
M.~Postma,
Phys.\ Rev.\ D {\bf 65}, 085035 (2002)

\bibitem[Randall,~Soljacic,~and Guth, 1996]{supernatural}
L.~Randall, M.~Soljacic and A.~H.~Guth,
Nucl.\ Phys.\ B {\bf 472}, 377 (1996)
[arXiv:hep-ph/9512439].

\bibitem[Randall and Thomas, 1995]{weakscale_infl}
L.~Randall and S.~Thomas,
Nucl.\ Phys.\ B {\bf 449}, 229 (1995)
[arXiv:hep-ph/9407248].

\bibitem[Rosen, 1968] {rosen}
G.~Rosen: J. Math. Phys. {\bf 9}, 996 (1968)
ibid. {\bf 9}, 999 (1968).


\bibitem[Rubakov and Shaposhnikov, 1996]{rs_review}
V.~A.~Rubakov and M.~E.~Shaposhnikov,
Usp.\ Fiz.\ Nauk {\bf 166}, 493 (1996)
[Phys.\ Usp.\  {\bf 39}, 461 (1996)]
[arXiv:hep-ph/9603208].

\bibitem[Rummukainen {\em et al.}, 1998]{EW_num3}
K.~Rummukainen, M.~Tsypin, K.~Kajantie, M.~Laine and M.~E.~Shaposhnikov,
Nucl.\ Phys.\ B {\bf 532}, 283 (1998).

\bibitem[Sakharov, 1967]{sakharov}
A. D. Sakharov, JETP Lett. {\bf 6}, 24 (1967).

\bibitem[Shaposhnikov, 1986]{ms86}
M.~E.~Shaposhnikov,
JETP Lett.\  {\bf 44}, 465 (1986)
[Pisma Zh.\ Eksp.\ Teor.\ Fiz.\  {\bf 44}, 364 (1986)].

\bibitem[Shaposhnikov, 1987]{ms87}
M.~E.~Shaposhnikov,
Nucl.\ Phys.\ B {\bf 287}, 757 (1987).

\bibitem[Smit and Tranberg, 2002]{smit}
J.~Smit and A.~Tranberg,
JHEP {\bf 0212}, 020 (2002)
[arXiv:hep-ph/0211243].

\bibitem[SNO, 2002] {sno}
SNO Collaboration: Q.~R.~Ahmad {\it et al.},
Phys.\ Rev.\ Lett.\  {\bf 89}, 011301 (2002)
[arXiv:nucl-ex/0204008].

\bibitem['t Hooft, 1976]{thooft}
G.~'t Hooft,
Phys.\ Rev.\ Lett.\  {\bf 37}, 8 (1976).

\bibitem[Weinberg, 1976]{lw2}
S.~Weinberg, Phys. Rev. Lett. {\bf 37}, 657 (1976).

\bibitem[Witten, 2002]{witten_GUT_2}
E.~Witten,
arXiv:hep-ph/0207124.


\end{thebibliography}

\end{document}